\documentclass[a4paper,11pt]{article}
\usepackage{jheppub} 
\usepackage{lineno}
\usepackage{mathtools}
\usepackage{braket}
\usepackage[T1]{fontenc}

\def\bal#1\eal{\begin{align}#1\end{align}}

\title{Time evolution of scalar condensate decay}

\author{
{\large Ayuki Kamada$^{1}$, Kodai Sakurai$^{2}$}
\\*[20pt]
{\it \normalsize 
$^1$Institute of Theoretical Physics, Faculty of Physics, University of Warsaw, \\
ul.~Pasteura 5, PL-02-093 Warsaw, Poland \\[5pt]
$^2$National Institute of Technology, Tsuruoka College, Tsuruoka, Yamagata 997-0842, Japan
} \\*[5pt]
}

\emailAdd{akamada@fuw.edu.pl, kodai.sakurai@tsuruoka-nct.ac.jp}

\abstract{
A scalar field, which oscillates coherently over the space and decays through production of daughter particles, plays an important role in cosmology. 
It was recently shown that the parametric-resonance and Feynman-diagrammatic approaches give the same decay rate for a Bosonic daughter particle at a large time duration.
We study how the parametric-resonance result approaches to this asymptotic value, by numerically following the time evolution of the phase density of the daughter particle.
We see a difference among different instability bands in the narrow-resonance regime.
The lowest-order instability band approaches to the Feynman-diagrammatic result in tens of scalar oscillation periods, while the high-order instability bands approach dependently also on a coupling between the scalar and daughter particles.
This would infer that the total decay width, to which the lowest-order instability band contributes most dominantly in the perturbation theory, approaches to the Feynman-diagrammatic result also independently of the coupling.
To confirm it, we consider a simpler and analytically solvable quantum mechanical model: Rabi models.
In the Bosonic Rabi model, the parametric-resonance and Feynman-diagrammatic results agree with each other for all the time.
On the other hand, in the Fermionic Rabi model, they disagree at a large time duration.
This difference would be attributed to Bose enhancement vs Pauli blocking.
}

\begin{document}
\maketitle
\flushbottom

\section{Introduction}\label{sec:intro}
When a macroscopic number of particles occupy the same one-particle states, a whole state (or condensation) would be well described by a ``classical'' field, following the classical equation of motion.
The decay of a condensation is quantum rather than classical, since it leads to production of daughter particles from vacuum.
One famous example is Schwinger effect~\cite{Schwinger:1951nm} (see Refs.~\cite{Kiefer:1991xy, Kiefer:1993fw} for the wave functional approach).
Electric field (constant both in time and in space) can be regarded as a condensation of photons, and leads to pair-production of electron and positron.
Another example, which we consider in this paper, is a coherently oscillating scalar field.
A scalar field (constant in space) oscillates in time and produce daughter particles~\cite{Dolgov:1989us, Traschen:1990sw} (see also, e.g., Refs.~\cite{Kofman:1994rk, Kofman:1997yn} for parametric resonances).
Such a scalar field plays an important role in cosmology.
A notable example is an inflaton field, which drives the inflation period and whose decay leads to reheating of the Universe (see, e.g., Ref.~\cite{Kolb:1990vq}).

There are two classes of approaches to compute the decay rate (per volume) of a scalar field. (Technically speaking, they are valid only when self-interaction of daughter particles is negligible and when scalar field can be treated as a background field.)
One is based on a produced number of daughter particles.
It is related with the decay of a scalar field in the wave functional approach~\cite{Yoshimura:1995gc}.
The produced number of daughter particles are evaluated from their mode functions, which evolve from a quantum seed through their equation of motion in the presence of an external scalar condensate.
We dub this approach as a parametric-resonance approach for a reason discussed below.
Another is based on an effective action for vacuum-to-vacuum transition of daughter particles in the presence of an external scalar condensate~\cite{Kamada:2025evo} (see also, e.g., Ref.~\cite{Matsumoto:2007rd} for the $S$-matrix approach, and Refs.~\cite{Morikawa:1986rp,Boyanovsky:1994me,Greiner:1996dx,Yokoyama:2004pf,Berera:2008ar,Mukaida:2013xxa,Wang:2022mvv,Ai:2023ahr} for non-equilibrium quantum field theory).
The decay of a scalar field is given by an imaginary part of the effective action.
Actual computation often needs to rely on perturbation theory.
The effective action is given by a sum of vacuum bubbles with all possible insertions of interaction with an external scalar condensate.
Its imaginary part is evaluated by the cutting rule.
We dub this approach as a Feynman-diagrammatic approach.

Recent Ref.~\cite{Kamada:2025evo} shows that these two approaches give the same result for a large time duration in a simple model with a cubic coupling (quadratic in daughter particles and linear in a scalar condensate).
In the parametric-resonance approach, the mode functions follow a so-called Mathieu equation, which admits growing solutions called parametric resonances.
The decay rate at a large time duration is related with the growth rate of the growing solutions.
The results at low orders in perturbation theory are found to be those of the Feynman-diagrammatic approach.
This equivalence sounds reasonable, since the parametric-resonance approach and Feynman-diagrammatic approach compute the same vacuum-to-vacuum transition in different ways.

On the other hand, generalizing this equivalence, say, to the following two cases is not obvious.
One is for a small time duration.
In the parametric-resonance approach, the decay rate can be computed for any time duration at least numerically.
The Feynman-diagrammatic approach often requires a large time duration to keep the Feynman rules simple and the same as for usual particle-physics computations.
It is not only interesting but also important for applications, to know how this (possibly) time-dependent decay rate approaches to the constant value, with which the Feynman-diagrammatic approach agrees. 
Another is for Fermionic daughter particles.
The Feynman-diagrammatic approach will not see a qualitative difference from Bosonic daughter particles.
Cutting rules apply to Fermionic loops in a similar way, though propagators are more complicated and loops acquire a minus sign.
On the other hand, the parametric-resonance approach will see a drastic change.
This is because Pauli blocking prevents a growing solution of the mode functions and thus there is no growth rate~\cite{Baacke:1998di, Greene:1998nh, Greene:2000ew} (see also, e.g., Ref.~\cite{Asaka:2010kv} for perturbative computation of the mode functions). 

In this paper, we study these questions in the following way.
In section~\ref{sec:calc}, after briefly reviewing the result of Ref.~\cite{Kamada:2025evo}, we solve the equation of motion perturbatively to find the mode function and the decay rate analytically. Then, we numerically solve the equation of motion and compare all the results.
To better understand these results, in section~\ref{sec:rabi} we introduce a Bosonic Rabi model in quantum mechanics, for which the decay rate is obtained analytically for any time duration.
We also compute the decay rate perturbatively by employing Feynman rules in quantum mechanics.
We discuss the results especially in comparison with a Fermionic Rabi model.
Section~\ref{sec:conclusions} is devoted to conclusions.

\section{Time evolution in the parametric resonance approach}\label{sec:calc}
\subsection{Review of Ref.~\cite{Kamada:2025evo}}
As in Ref.~\cite{Kamada:2025evo}, we consider the following toy model (Lagrangian) with two real scalars:
\bal
{\cal L} = \frac{1}{2} (\partial \varphi)^2 + \frac{1}{2} (\partial \chi)^2 - \frac{1}{2} m_\varphi^2 \varphi^2 - \frac{1}{2} m_\chi^2 \chi^2 - \frac{1}{2} \mu \varphi \chi^2 \,.
\eal
The last term is an interaction between the two scalars.
Without the interaction, $\varphi$ admits the oscillating (in a spatially coherent manner) solution as
\bal
\varphi = A_\varphi \sin (m_\varphi t) \,.
\eal
(We have changed $\cos \to \sin$ from Ref.~\cite{Kamada:2025evo} for our convenience. It involves the change in an oscillation phase, but does not affect the results.)
In the presence of this background solution, the equation of motion of $\chi$ in the Fourier space is given by
\bal \label{eq:eom}
\frac{d^2}{dt^2} \chi_{k}(t) + E_k^2 (t) \chi_{k}(t) = 0
\eal
with $E_k^2 (t) = k^2 + m_\chi^2 + \mu A_\varphi \sin (m_\varphi t) = \omega_k^2 + \mu A_\varphi \sin (m_\varphi t)$.

This is known as the Mathieu equation and admits the growing solution $\chi_{k}(t) \propto \exp(\lambda (k) m_\varphi t / 2)$ with a dimensionless growth rate $\lambda (k)$, in a certain range of the momentum. We focus on the so-called narrow resonance regime and thus these instability bands are separated into $n_\varphi$ contributions:
\bal
k_{n_\varphi} = \frac{n_\varphi m_\varphi}{2} \beta_{n_\varphi} \sqrt{1 + \frac{\epsilon_{n_\varphi}}{(n_\varphi \beta_\varphi)^2}} \,, \quad \beta_{n_\varphi} = \sqrt{1 - \left(\frac{2 m_\chi}{n_\varphi m_\varphi} \right)^2} \,,
\eal
where a dimensionless energy is $\epsilon_{n_\varphi} = 2 E_k (t) / m_\varphi - n_\varphi$.
The $n_\varphi = 1$ contribution can be interpreted as decay of $\varphi \to \chi \chi $:
\bal
\lambda_{n_\varphi = 1}(k) = \frac{1}{2} \sqrt{\theta^2 - \epsilon_{n_\varphi = 1}^2}
\eal
in the leading order of $\theta = 2 \mu A_\varphi / m_\varphi^2$.
The growth rate is $\lambda_{n_\varphi = 1} = \theta / 2$ at the peak $\epsilon_{n_\varphi = 1} = 0$ and the instability band around the peak has the width of $\pm \theta$.
The $n_\varphi = 2$ contribution can be interpreted as annihilation of $\varphi \varphi \to \chi \chi $:
\bal
\lambda_{n_\varphi = 2}(k) = \frac{1}{4} \sqrt{\frac{\theta^4}{16} - \left(\epsilon_{n_\varphi = 2} - \frac{\theta^2}{6} \right)^2}
\label{eq:k_res}
\eal
in the leading order of $\theta$.
The growth rate is $\lambda_{n_\varphi = 1} = \theta^2 / 16$ at the peak $\epsilon_{n_\varphi = 2} = \theta^2 / 6$ and the instability band around the peak has the width of $\pm \theta^2 / 4$.

We analyze the decay (damping) rate of the oscillating solution in two different ways: one is based on solving the Schrodinger equation (for a wave functional) and relating the vacuum-to-vacuum amplitude (for $\chi$) to the growing solution of the equation of motion; the other is based on relating the vacuum-to-vacuum amplitude to the vacuum bubbles (of $\chi$) and computing the vacuum bubbles with Feynman rules in the Fourier space.


More specifically, in the parametric resonance approach, the vacuum-to-vacuum amplitude for any (not necessarily large) duration is given by
\bal
|_{\rm out}\langle 0 | 0 \rangle_{\rm in}| = \prod_{\vec{k}} \left[ 1 + f_{k} (T) \right]^{-1/4} \,,
\eal
where
\bal \label{eq:fchi}
f_{k} (t) = \frac{1}{2 E_k(t)} \left[ \left| \frac{d}{dt} \chi_k(t) \right|^2 + E_k^2(t) |\chi_k(t)|^2 \right] - \frac{1}{2}
\eal
is interpreted as a phase-space distribution of produced $\chi$ with the initial conditions of
\bal
\chi_{k}(t=0) = \frac{1}{\sqrt{2 \omega_k}} \,, \quad \frac{d}{dt} \chi_{k}(t=0) = - i \omega_k \chi_{k}(t=0)\,.
\eal
[We have changed the normalization of $\chi_k$ by $\sqrt{2 \pi}$ also $ i \omega_k \to -i \omega_k$ (complex conjugate of $\chi_k$) from Ref.~\cite{Kamada:2025evo} for our convenience.]
This gives us a (volumetric) decay rate of the oscillating scalar $\varphi$, which is defined by
\bal
|_{\rm out}\langle 0 | 0 \rangle_{\rm in}| = \exp \left( - \Gamma (T) L^3 T / 2 \right) \,,
\eal
with a volume $L^3$, as
\bal
&\Gamma^P (T) = \frac{1}{2 T} \int \frac{d^3 k}{(2 \pi)^3} \ln\left[ 1 + f_{k} (T) \right] =  \frac{1}{2 T} \frac{1}{2 \pi^2} \int_{m_\chi}^{\infty} d\omega_k \omega_k^2 \beta_k \ln\left[ 1 + f_{k} (T) \right] \notag \,, \\
& \beta_k=\sqrt{1 - \frac{m_\chi^2}{\omega_k^2}} \,,
\label{eq:GammaP}
\eal
in the parametric-resonance approach.

At a large $T$, namely, after the growing modes get dominant, the decay rate is approximated as
\bal
&\Gamma^P (T) \approx \frac{m_\varphi}{2} \frac{1}{2 \pi^2} \int_{m_\chi}^{\infty} d\omega_k \omega_k^2 \beta_k \lambda(k) \,.
\eal
By decomposing the $k$ integral into each instability band, Ref.~\cite{Kamada:2025evo} finds
\bal
&\Gamma_{n_\varphi=1}^P = \frac{m_\varphi^4}{16 \pi} \beta_1 \left( \frac{\mu A_\varphi}{2 m_\varphi^2} \right)^2 - \frac{m_\varphi^4}{32 \pi} \beta_1 \left( 2 + \frac{3}{\beta_1^2} + \frac{1}{\beta_1^4} \right) \left( \frac{\mu A_\varphi}{2 m_\varphi^2} \right)^4 + \dots \,, \\
&\Gamma_{n_\varphi=2}^P = \frac{m_\varphi^4}{16 \pi} \beta_2 \left( \frac{\mu A_\varphi}{2 m_\varphi^2} \right)^4 + \dots \,, 
\eal
in the relevant leading orders of the coupling $\mu$. Note that they do not depend on $T$ and thus can be regarded as an asymptotic value.
In the following we examine how $\Gamma^P (T)$ approaches to these asymptotic values.
Ref.~\cite{Kamada:2025evo} finds the same result in the Feynman-diagrammatic approach $\Gamma^F (T)$ for a large $T$.
This large $T$ is to enforce energy-momentum conservation by the delta functions at each vertex in the Fourier space.
The delta function is expected to be a good approximation when the integration interval in the real space is bigger than the inverse of the typical energy scale (from dimensional analysis): namely, $T \gg 1 / m_\varphi$ in our case independently of $\mu A_\varphi$ unlike growth factor of the parametric resonance.

\subsection{Perturbative computation of the phase-space distribution}
Before solving the full equation of motion numerically, we introduce a perturbative method to solve the equation of motion for $\chi$. The idea is to construct a formal solution by using Green function as follows:
\bal
\chi_k(t) = \frac{1}{2 \omega_k} e^{-i \omega_k t} + \int^\infty_0 dt' G(t-t') (-\mu A_\varphi) \sin (m_\varphi t') \chi_k(t') \,, \quad \frac{d^2}{dt^2} G(t-t') + \omega_k^2 G(t-t') = \delta(t-t') \,.
\eal
By remembering a general construction of Green function in terms of  Wronskian for an ordinary differential equation, one finds the (retarded) Green function as
\bal
G(t-t') = \frac{1}{2 i \omega_k} \left[e^{i \omega_k (t - t')} - e^{-i \omega_k (t - t')} \right] \theta(t-t').
\eal
Therefore, the formal solution is given by
\bal
\chi_k(t) = \frac{1}{2 \omega_k} e^{-i \omega_k t} + \int^t_0 dt' \frac{1}{2 i \omega_k} \left[e^{i \omega_k (t - t')} - e^{-i \omega_k (t - t')} \right] (-\mu A_\varphi) \sin (m_\varphi t') \chi_k(t') \,.
\eal
One can check that this satisfies the initial conditions.

Then, one can solve the formal solution iteratively (like the Born series of the Lippmann-Schwinger equation in quantum mechanics):
\bal
&\chi_k(t) = \chi^{(0)}_k(t) + \chi^{(1)}_k(t) + \dots \,, \quad \chi^{(0)}_k(t) = \frac{1}{2 \omega_k} e^{-i \omega_k t} \,, \notag \\
&\chi^{(n+1)}_k(t) = \int^t_0 dt' \frac{1}{2 i \omega_k} \left[e^{i \omega_k (t - t')} - e^{-i \omega_k (t - t')} \right] (-\mu A_\varphi) \sin (m_\varphi t') \chi^{(n)}_k(t') \,.
\eal
Here are the low-order solutions:
\bal
&\chi^{(1)}_k(t) = - i \frac{\mu A_\varphi}{8\sqrt{2} m_\varphi \omega _k^{3/2} \left[\omega_k^2 - (m_\varphi/2)^2 \right]} \Big[-2 \omega_k \left(\omega_k + m_\varphi/2 \right) e^{-i \left(\omega_k - m_\varphi \right) t} 
- 2 \omega_k \left(\omega_k - m_\varphi/2 \right) e^{-i \left(\omega_k + m_\varphi \right) t} \notag\\    
&\qquad \qquad + 4 \left[\omega_k^2 - (m_\varphi/2)^2 \right] e^{-i \omega_k t} + m_\varphi^2 e^{i \omega_k t} \Big] \,, \\
&\chi^{(2)}_k(t) =
-\frac{(\mu A_\varphi)^2}{128 \sqrt{2} \omega_k^{5/2} m_\varphi^2 \left[\omega_k^2 - (m_\varphi/2)^2 \right]^2}
\Bigg[
-16 \omega_k \left(\omega_k - m_\varphi/2 \right) \left(\omega_k + m_\varphi/2 \right)^2 e^{- i \left(\omega_k - m_\varphi \right) t }\notag \\ 
&\qquad \qquad + 4 \omega_k m_\varphi^2 \left(\omega_k - m_\varphi / 2 \right) e^{i \left(\omega_k + m_\varphi \right) t} + 4 \omega_k m_\varphi^2 \left(\omega_k + m_\varphi/2 \right) e^{i \left(\omega_k - m_\varphi \right) t} \notag\\
&\qquad \qquad +4 \frac{\omega_k^2 \left(\omega_k - m_\varphi/2 \right)^2 \left(\omega_k + m_\varphi/2 \right)}{\omega_k + m_\varphi} e^{-i \left(\omega_k + 2m_\varphi \right) t} - 16 \omega_k \left(\omega_k - m_\varphi/2 \right)^2 \left(\omega_k + m_\varphi/2 \right) e^{-i \left(\omega_k + m_\varphi \right) t} \notag \\
&\qquad \qquad + 4\frac{\omega_k^2 \left(\omega_k - m_\varphi/2 \right) \left(\omega_k + m_\varphi/2 \right)^2}{\omega_k - m_\varphi} e^{-i \left(\omega_k - 2 m_\varphi \right) t}
-8 \frac{m_\varphi^2\left[\omega_k^2 - (m_\varphi/2)^2 \right] \left[\omega_k^2 + m_\varphi^2/2 \right]}{\omega_k^2 - m_\varphi^2} e^{i \omega_k t} \notag \\
&\qquad \qquad + \left\{ \left(24 \omega_k^4 - 18 m_\varphi^2 \omega_k^2 + m_\varphi^4 \right) - 8 i t \omega_k m_\varphi^2 \left[\omega_k^2 - (m_\varphi/2)^2 \right] \right\} e^{-i \omega_k t} \Bigg] \,.
\eal
Though we also compute $\chi^{(3)}_k(t)$ and $\chi^{(4)}_k(t)$, their expressions are too long to show here.

By using them, one can also compute the distribution function in a perturbative way:
\bal
f_k(t) = f^{(0)}_k(t) + f^{(1)}_k(t) + \dots \,.
\eal
In the following, we focus on the time being multiples of scalar oscillation period $t_N = 2 \pi N / m_\varphi$ with $N=0, 1, \dots$, so that we do not need to expand $E_k(t)$.
Here are the low-order contributions:
\bal
&f^{(0)}_k (t_N) = f^{(1)}_k (t_N) = f^{(3)}_k (t_N) = 0 \,,  \\
&f_k^{(2)}(t_N) = \left( \frac{\mu A_\varphi}{2 m_\varphi^2} \right)^2 \frac{m_\varphi^6}{4 \omega_k^2 \left[\omega_k^2 - (m_\varphi/2)^2 \right]^2} \sin^2(\omega_k t_N) \,,  \\
&f_k^{(4)}(t_N) = - \left( \frac{\mu A_\varphi}{2 m_\varphi^2} \right)^4 \frac{m_\varphi^{10}}{1024 {\omega_k^4} \left[\omega_k^2 - (m_\varphi/2)^2 \right]^4 \left(\omega_k^2 - m_\varphi^2 \right)^2 \left[\omega_k^2 - (3 m_\varphi/2)^2 \right]}  \notag \\
&\qquad \qquad ~ \times \Big[128 \omega_k t_N \left[\omega_k^2 - (m_\varphi/2)^2 \right] \left(\omega_k^2 - m_\varphi^2 \right)^2 \left[\omega_k^2 - (3 m_\varphi/2)^2 \right] \sin \left(\omega_k t_N \right) \cos \left(\omega_k t_N \right) \notag \\
&\qquad \qquad ~ + m_\varphi^2 \left(1664 \omega_k^6 - 1888 m_\varphi^2 \omega_k^4 + 620 m_\varphi^4 \omega_k^2 + 9 m_\varphi^6 \right) \sin^2 \left(\omega_k t_N \right) \Big] \,.
\eal
Vanishing odd orders is expected from the fact that the sign of $\mu A_{\varphi}$ is altered by the change of the oscillation phase and thus not physical.

$f_k^{(2)}(t_N)$ and $f_k^{(4)}(t_N)$ have possible resonances at $\omega_k = m_\varphi/2$ and $\omega_k = m_\varphi/2, m_\varphi, 3 m_\varphi/2$, respectively.
Expansion around these resonances is as follows:
\bal
&f_k^{(2)}(t_N) = \left( \frac{\mu A_\varphi}{2 m_\varphi^2} \right)^2 m_\varphi^2 t_N^2 \Bigg[1 - \frac{6}{m_\varphi} \left(\omega_k - m_\varphi/2 \right) - \frac{{m_\varphi^2}t_N^2 - 69}{3 m_\varphi^2} \left(\omega_k - m_\varphi/2 \right)^2 + \dots \Bigg] \,, \\
&f_k^{(4)}(t_N) = \left( \frac{\mu A_\varphi}{2 m_\varphi^2} \right)^4 \frac{m_\varphi^2 t_N^2 (2 m_\varphi^2 t_N^2 - 9)}{6} \Bigg [1 - \frac{3 \left(12 m_\varphi^2 t_N^2 - 79 \right)}{2 m_\varphi (2 m_\varphi^2 t_N^2 - 9)} \left(\omega_k - m_\varphi/2 \right) \notag \\
&\qquad \qquad - \frac{32 m_\varphi^4 t_N^4 - 5940 m_\varphi^2 t_N^2 + 33995}{60 m_\varphi^2 (2 m_\varphi^2 t_N^2 - 9)} \left(\omega_k - m_\varphi/2 \right)^2 + \dots \Bigg] \,, \\
&f_k^{(4)}(t_N) = \left( \frac{\mu A_\varphi}{2 m_\varphi^2} \right)^4 \frac{m_\varphi^2 t_N^2}{2} \Bigg[1 - \frac{167}{27 m_\varphi} \left(\omega_k - m_\varphi \right) - \frac{540 m_\varphi^2 t_N^2 - 53207}{1620 m_\varphi^2} \left(\omega_k - m_\varphi \right)^2 + \dots \Bigg]\,, \\
&f_k^{(4)}(t_N) = \left( \frac{\mu A_\varphi}{2 m_\varphi^2} \right)^4 \frac{7 m_\varphi^2 t_N^2}{216}
\Bigg[- \frac{1}{m_\varphi}\left(\omega _k- 3 m_\varphi/2 \right) + \frac{4523}{525 m_\varphi^2} \left(\omega_k - 3 m_\varphi/2 \right)^2 + \dots \Bigg] \,.
\eal

Resonances of $f_k^{(2)}(t_N)$ at $\omega_k = m_\varphi/2$ and of $f_k^{(4)}(t_N)$ at $\omega_k = m_\varphi$ have peaks evolving in proportion to $t_N^2$ and widths evolving in proportion to $1 / t_N$ at a large $t_N$.
This factor of $\sin^2(\omega t_N) / \omega^2$ is familiar from the discussion about the Fermi golden rule in standard textbooks (e.g., Ref.~\cite{Sakurai:2011zz}) and is supposed to be approximated by $\pi \delta(\omega) t_N$ at a large $t_N$.
\bal
f_{n_\varphi=1, k}^{(2)}(t_N) \approx \left( \frac{\mu A_\varphi}{2 m_\varphi^2} \right)^2 \pi m_\varphi^2 t_N \delta(\omega_k - m_\varphi/2) \,, \quad
f_{n_\varphi=2, k}^{(4)}(t_N) \approx \left( \frac{\mu A_\varphi}{2 m_\varphi^2} \right)^4 \frac{\pi}{4} m_\varphi^2  t_N \delta(\omega_k - m_\varphi)
\eal
can be interpreted as decay of $\varphi \to \chi \chi$ and annihilation of $\varphi \varphi \to \chi \chi$, respectively.
\bal
\Gamma^P (t_N) = \frac{1}{2 t_N} \frac{1}{2 \pi^2} \int d\omega_k \omega_k^2 \beta_k \left[ f^{(2)}_k (t_N) + \left( f^{(4)}_{k} (t_N) - \frac{1}{2} \left[ f^{(2)}_k (t_N) \right]^2 \right) + \dots \right]
\label{eq:GammaP_exp}
\eal
is approximated by
\bal
\Gamma_{n_\varphi=1}^P (t_N) \approx \frac{m_\varphi^4}{16 \pi} \beta_1 \left( \frac{\mu A_\varphi}{2 m_\varphi^2} \right)^2 \,, \quad \Gamma_{n_\varphi=2}^P (t_N) \approx \frac{m_\varphi^4}{16 \pi} \beta_2 \left( \frac{\mu A_\varphi}{2 m_\varphi^2} \right)^4 \,,
\eal
at a large $t_N$.
They reproduce the corresponding-order results of the Feynman-diagrammatic approach.

A resonance of $f_k^{(4)}(t_N)$ at $\omega_k = m_\varphi/2$ has a peak evolving in proportion to $t_N^4$ and widths evolving in proportion to $1 / t_N$ at a large $t_N$.
As a result, $f^{(4)}_{n_\varphi=1, k} (t_N) - \left[ f^{(2)}_{n_\varphi=1, k} (t_N) \right]^2 / 2$ contribution to $\Gamma_{n_\varphi=1}^P (t_N)$ evolves in proportion to $t_N^2$.
However, this time dependence does not contradict with the agreement between the parametric-resonance and Feynman-diagrammatic results (constant with time) at a large $t_N$~\cite{Kamada:2025evo}.
This time dependence means that even if $2 \mu A_\varphi / m_\varphi^2$ is small, higher-order contributions become more significant at a large $t_N$ and thus a perturbative computation breaks down.
Actually, this breakdown occurs around $2/m_\varphi/(2 \mu A_\varphi / m_\varphi^2)$ as expected from the growth rate of the parametric resonance [see the discussion around Eq.~\eqref{eq:k_res}].
A possible resonance of $f_k^{(4)}(t_N)$ at $\omega_k = 3m_\varphi/2$ does not have an actual peak.

\subsection{Numerical computation of the phase-space distribution} \label{sec:numerical}
In this section, we discuss the numerical non-perturbative computation of the phase-space distribution of $\chi$.
In numerical computations, we take units of $m_\varphi = 2$. We fix $m_\chi = 0.5$ but consider $\mu A_\varphi = 0.1,\, 0.3,\, 0.8$.
We discretize the momentum space and solve the equation of motion [Eq.~\eqref{eq:eom}] for each momentum.
Once $\chi_k$ is obtained, the distribution function $f_k$ for each momentum can be computed from Eq.~\eqref{eq:fchi}.
Parametric resonance is expected to take place around the peak momentum of $k^{\rm res}_{n_\varphi} = n_\varphi \beta_\varphi$ with the width of order of $\delta k^{\rm res}_{n_\varphi} = (2 \mu A_\varphi / m_\varphi^2)^{n_\varphi}$ [see the discussion around Eq.~\eqref{eq:k_res}].
Therefore, we take the momentum region for the $n_{\varphi}$ contribution sufficiently large around the peak, and furthermore split it into the sub-regions to resolve resonances as follows:
\bal
k^{\rm ini}_{n_\varphi} < k < k^{\rm res}_{n_\varphi} - M \delta k^{\rm res}_{n_\varphi};\ k^{\rm res}_{n_\varphi} - M \delta k^{\rm res}_{n_\varphi} < k < k^{\rm res}_{n_\varphi} + M \delta k^{\rm res}_{n_\varphi};\ 
k^{\rm res}_{n_\varphi} + M \delta k^{\rm res}_{n_\varphi} < k < k^{\rm end}_{n_\varphi} .
\eal
Each momentum division is divided into 300 (6,000) grid points for $n_\varphi=1\, (2)$. 
In the numerical evaluations, we take $ k^{\rm ini}_{n_\varphi=1}=0$ and $k^{\rm end}_{n_\varphi=1}=1.6$ for the $n_\varphi=1$ case, while $k^{\rm ini}_{n_\varphi=2}=1.72$ and $k^{\rm end}_{n_\varphi=2}=2.3$ for the $n_\varphi=2 $ case.
A fudge factor $M$ is taken to be 10 in the case of $n_\varphi=2$ and $\mu A_\varphi=0.1$, otherwise we take $M$=2.
We check that modestly varying these values does not change the results.

\begin{figure}
    \centering
    \includegraphics[width=0.49\linewidth]{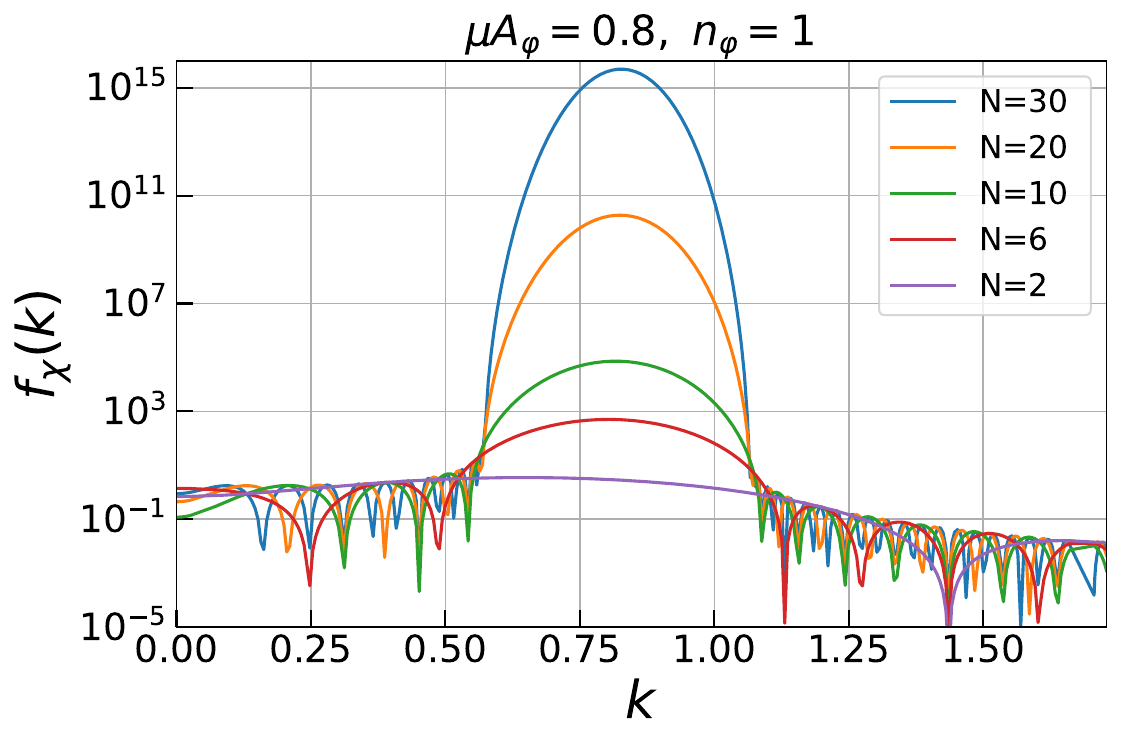}
    \includegraphics[width=0.49\linewidth]{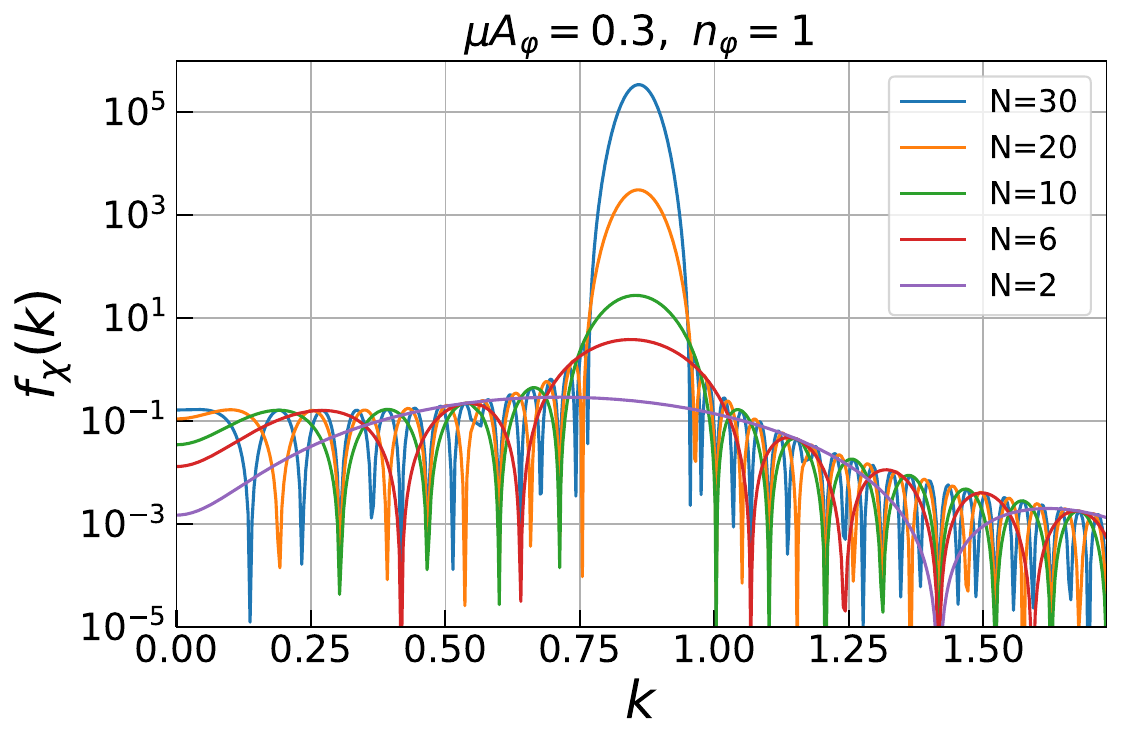}
    \includegraphics[width=0.49\linewidth]{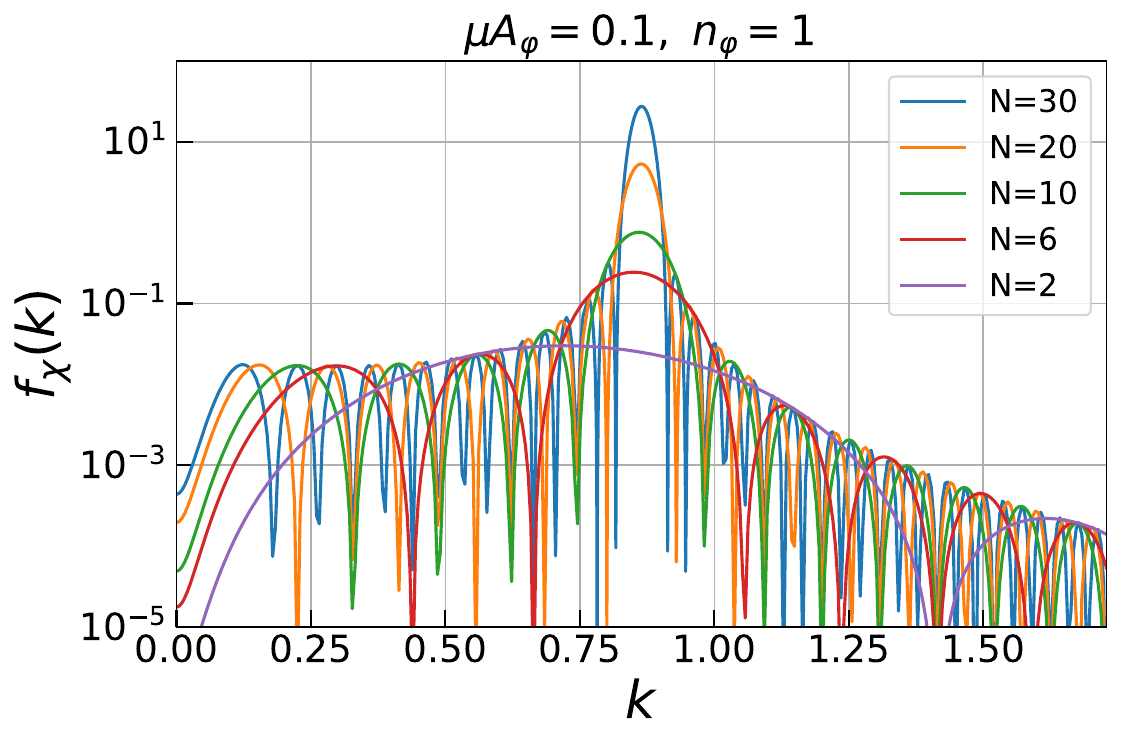}
    \caption{Evolution of the phase-space distribution for $n_\varphi=1$.}
    \label{fig:f_chi:nphi=1}
\end{figure}
\begin{figure}
    \centering
    \includegraphics[width=0.49\linewidth]{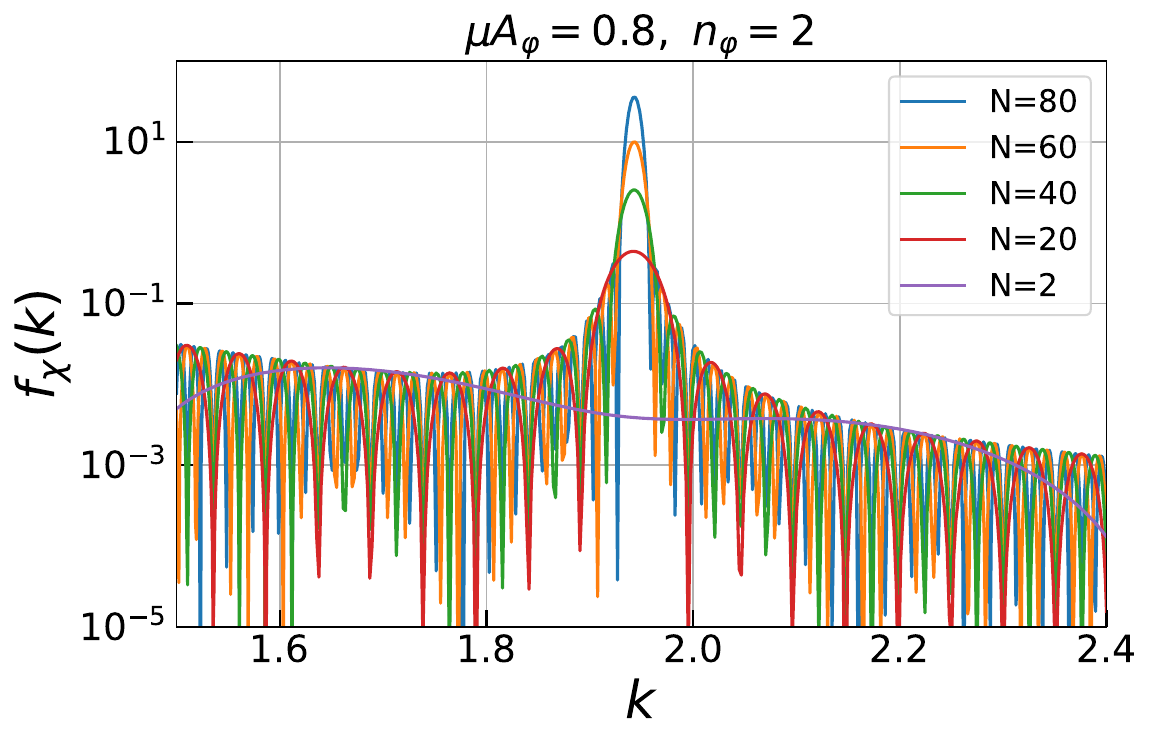}
    \includegraphics[width=0.49\linewidth]{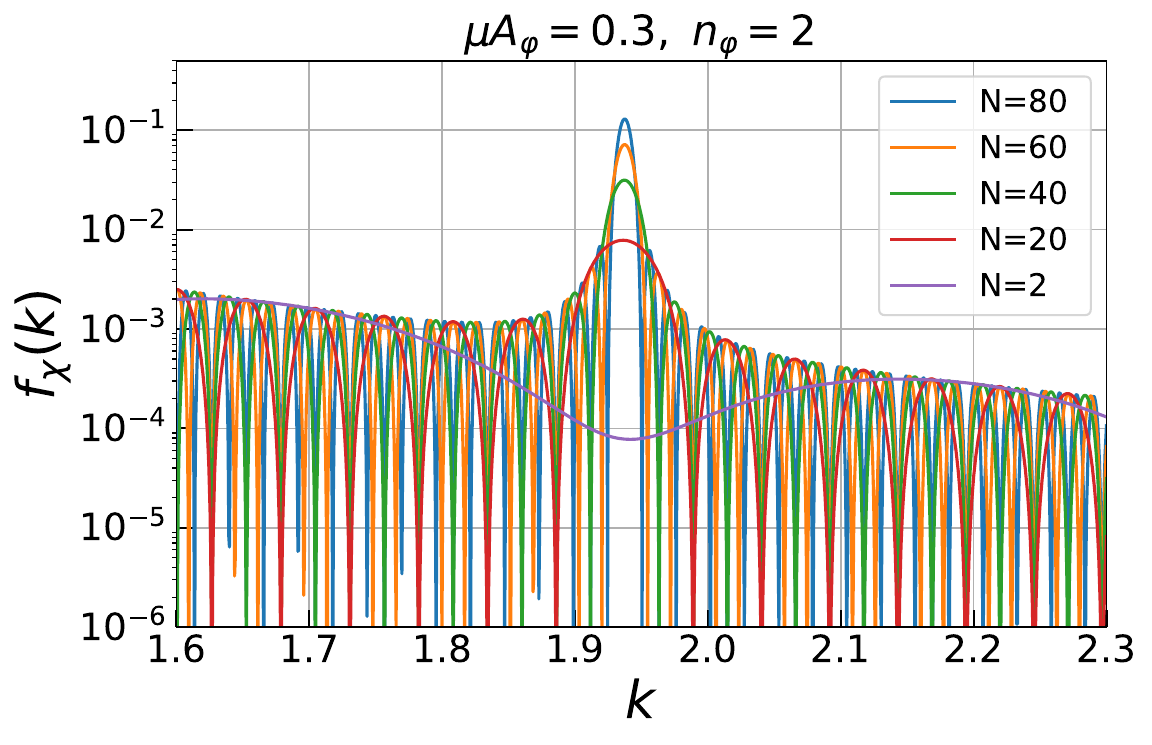}
    \includegraphics[width=0.49\linewidth]{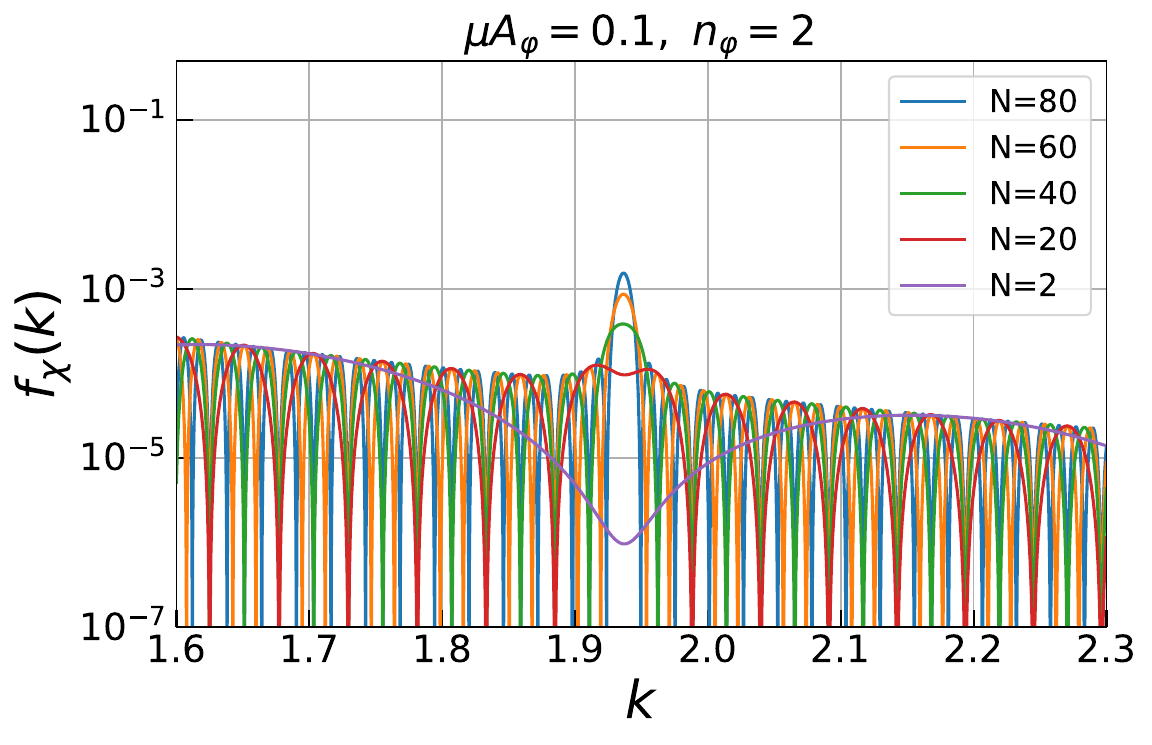}
    \caption{Evolution of the phase-space distribution for $n_\varphi=2$.}
    \label{fig:f_chi:nphi=2}
\end{figure}

\begin{figure}
    \centering
    \includegraphics[width=0.48\linewidth]{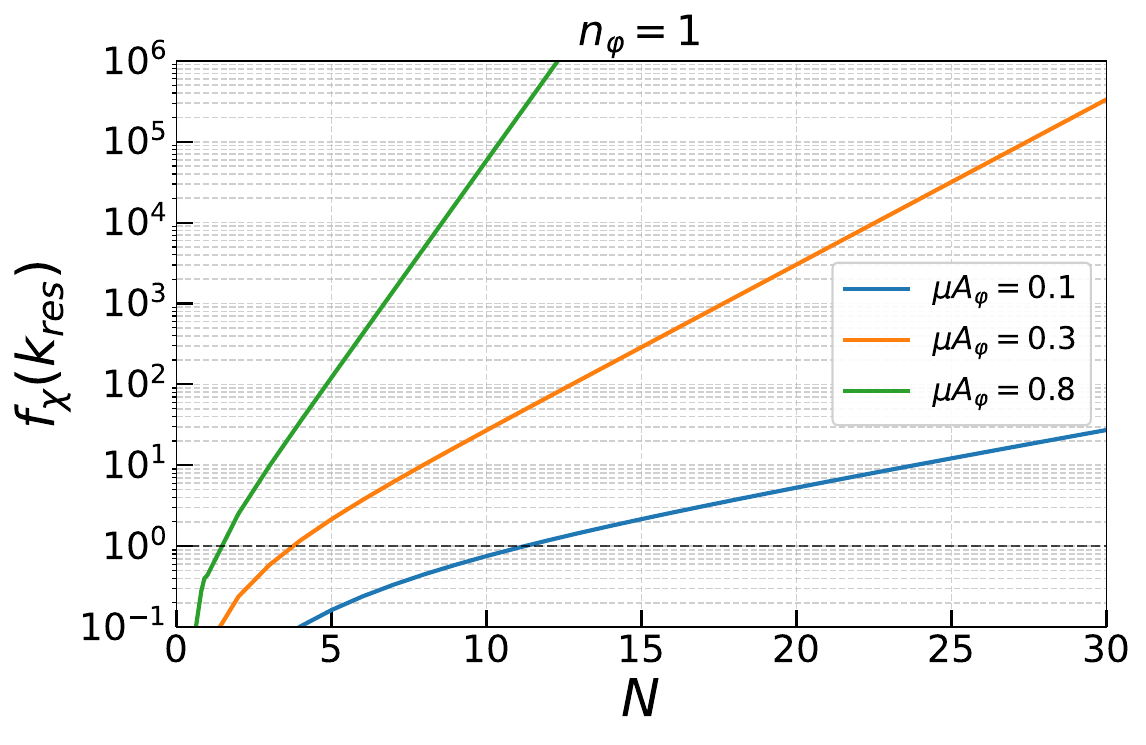}
    \includegraphics[width=0.48\linewidth]{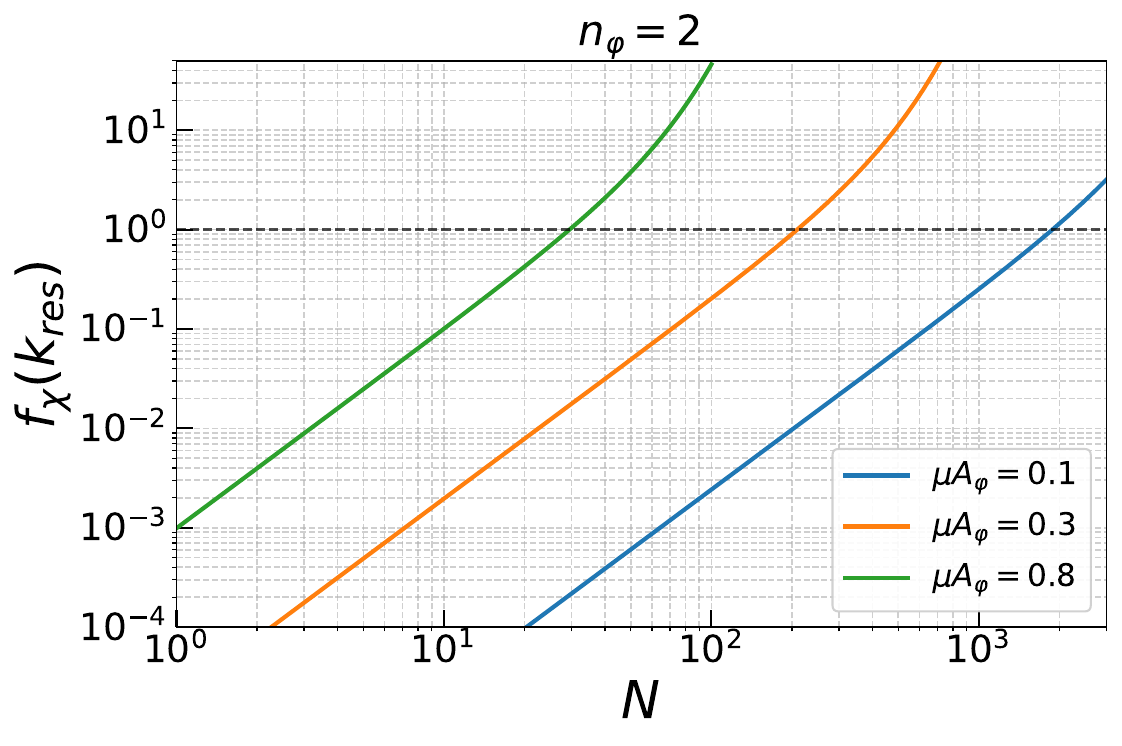}
    \caption{Evolution of the peak value of the phase-space distribution as a function of $N$. Note that the horizontal axes are linear for $n_\varphi = 1$ (left) and logarithmic for $n_\varphi = 2$ (right). {The black dashed horizontal lines correspond to a unity of the peak value.}}
    \label{fig:peak_value}
\end{figure}

The numerical evolution of the phase-space distribution $f_k$ are presented in Figs.~\ref{fig:f_chi:nphi=1} and \ref{fig:f_chi:nphi=2} for $n_\varphi=1$ and $n_\varphi=2$, respectively.
As expected from parametric resonance, the phase-space distribution exhibits a peak around $k^{\rm res}_{n_\varphi = 1} = 0.87 \, (1.94)$ for $n_\varphi=1 \, (2)$ at least for $N > 5$.
A width is proportional to $\mu A_\varphi$ for $n_\varphi=1$ as expected, while it is constant with $\mu A_\varphi$ for $n_\varphi=2$ (expected to be proportional to $(\mu A_\varphi)^2$).
This can be understood by examining if the peak value already experiences exponential growth or not yet.
To make it clear, we present the evolution of the peak value of the phase-space distribution in Fig.~\ref{fig:peak_value}.
The exponential growth starts roughly when the peak value exceeds unity.
Corresponding $N$ depends on $\mu A_\varphi$: roughly proportional to $1/(\mu A_\varphi)$ for $n_\varphi=1$ and to $1/(\mu A_\varphi)^2$ for $n_\varphi=2$, as expected from the growth rate of parametric resonance [see the discussion around Eq.~\eqref{eq:k_res}].
On the other hand, in the range of $N$ plotted in Fig.~\ref{fig:f_chi:nphi=2}, the peak is yet to grow exponentially.
This would be why the width does not follow the expected scaling yet.
We would not pursue the width for $n_\varphi=2$ further, since it is not our main objective and requires a finer resolution and a longer integration time.

\begin{figure}
    \centering
    \includegraphics[width=1.\linewidth]{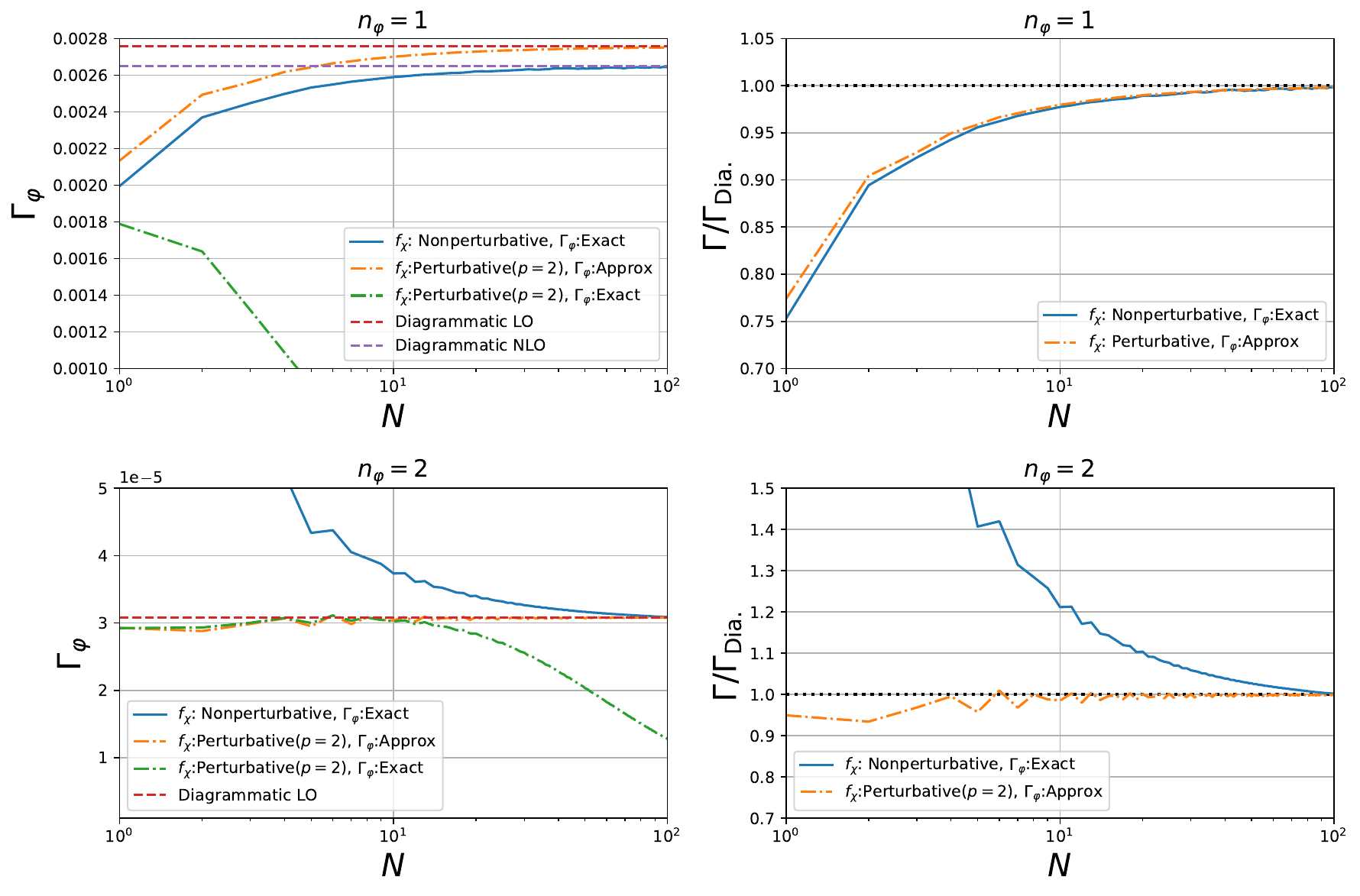}
    \caption{Decay rate as a function of $N$. Input parameters are taken as $\mu A_\varphi =0.8$. For $n_\varphi=1$, the ratio $\Gamma/\Gamma_{\rm Dia.}$ is defined by normalizing the non-perturbative (perturbative) result to the diagrammatic prediction up to NLO (LO). For $n_\varphi=2$, the same diagrammatic prediction at LO is used as the reference for both the non-perturbative and perturbative results. }
    \label{fig:muphiew0p8}
\end{figure}

\begin{figure}
    \centering
    \includegraphics[width=1.\linewidth]{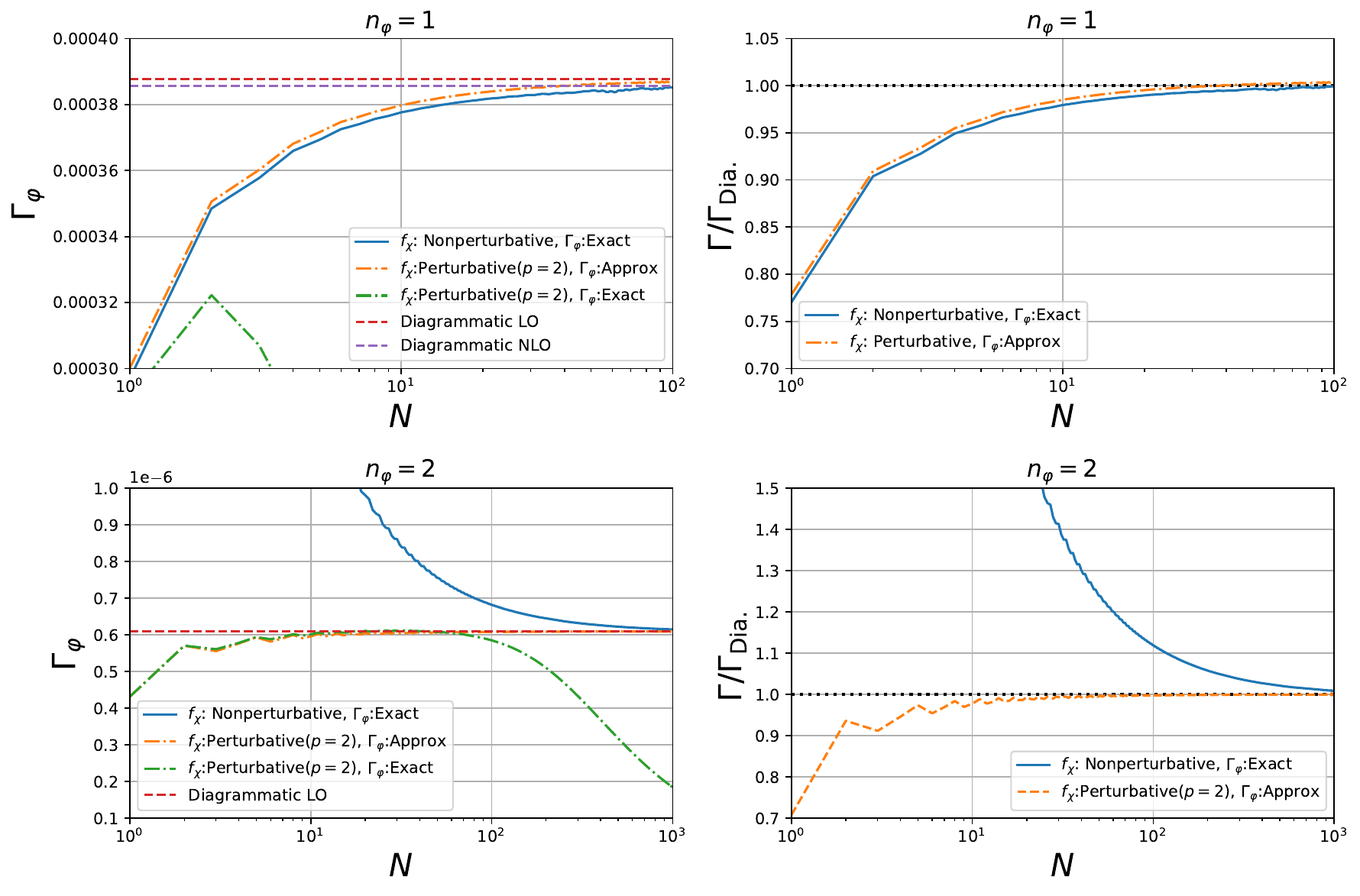}
    \caption{Decay rate as a function of $N$. Input parameters are taken as $\mu A_\varphi =0.3$. $\Gamma/\Gamma_{\rm Dia.}$ is defined in the same way as in Fig.~\ref{fig:muphiew0p8}. }
    \label{fig:muphiew0p3}
\end{figure}

\begin{figure}
    \centering
    \includegraphics[width=1.\linewidth]{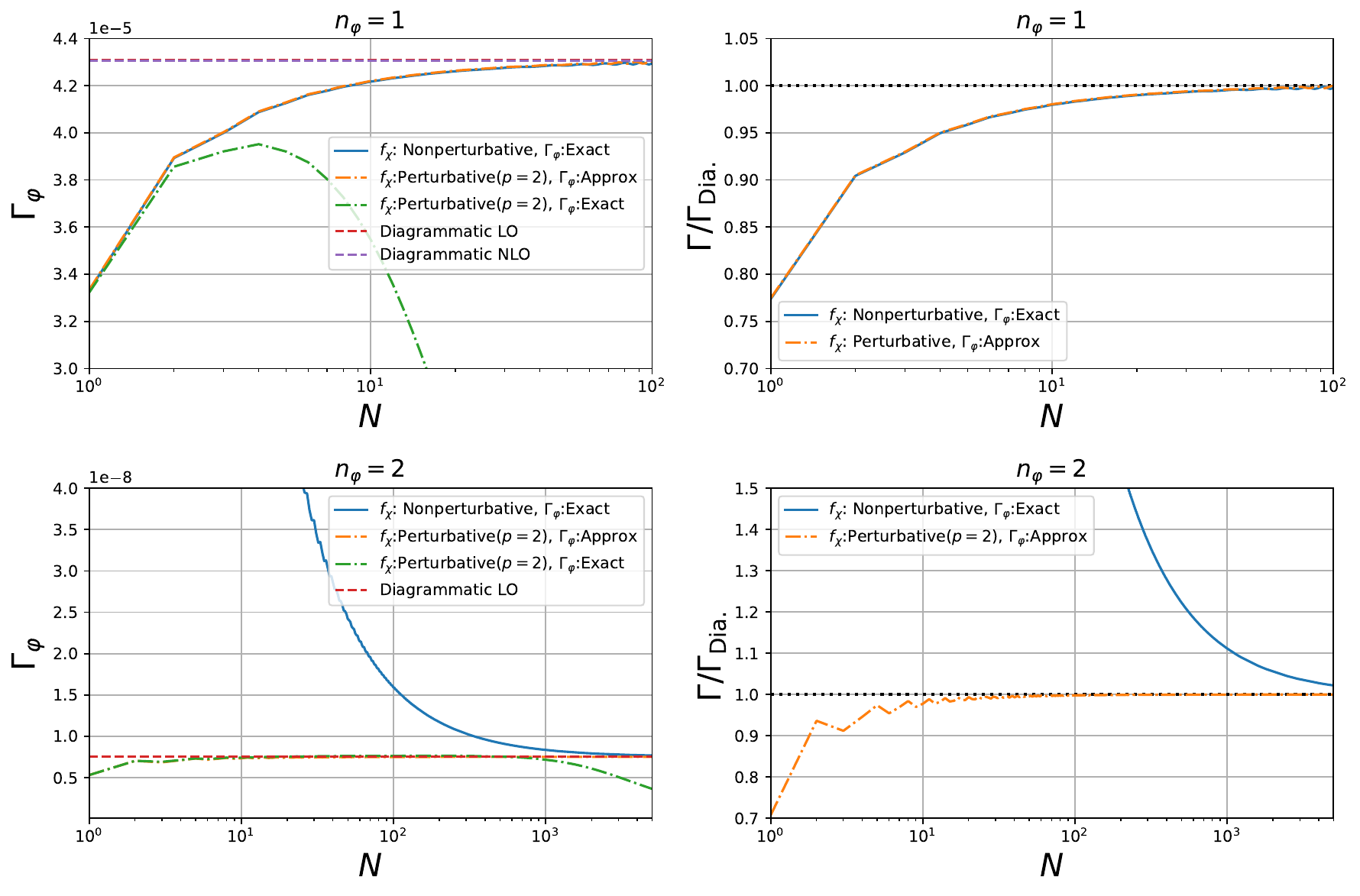}
    \caption{Decay rate as a function of $N$. Input parameters are taken as $\mu A_\varphi =0.1$ in units of $m_\varphi =2$. The $\Gamma/\Gamma_{\rm Dia.}$ is defined in the same way as in Fig.~\ref{fig:muphiew0p8}.  }
    \label{fig:muphiew0p1}
\end{figure}

From the obtained phase-space distribution, one can compute the decay rate as in Eq.~\eqref{eq:GammaP}.
The results for $\mu A_\varphi = 0.8$ are shown in Fig.~\ref{fig:muphiew0p8}.
For comparison, we also show the results obtained from the Feynman-diagrammatic calculation. 
Both the leading-order (LO) and up to the next leading-order (NLO) results are presented by the red and purple dashed lines, respectively, for $n_\varphi=1$ (top panels).
Only the LO results are presented for $n_\varphi=2$ (bottom panels), since the NLO  is not distinguishable from the LO.
In addition, the results obtained by perturbatively solving the phase-space distribution are presented.
With the LO phase-space distribution, we show both the result obtained from the LO contribution of $\Gamma_\phi$ in Eq.~\eqref{eq:GammaP_exp} (labeled ``$\Gamma_\phi$: Approx'', shown by the orange dot-dashed curve) and the result obtained using the full expression for $\Gamma_\phi$ in Eq.~\eqref{eq:GammaP} (labeled ``$\Gamma_\phi$: Exact'', shown by the green dot-dashed curve).

For the case of $n_\varphi=1$, we find that the non-perturbative result approaches the diagrammatic NLO prediction and agrees with it for $N > 100$.
Furthermore, even for relatively small values of $N$, the non-perturbative and diagrammatic results agree within approximately $10\%$.
The perturbative result based on the expanded expression of $\Gamma_\varphi$ approaches the diagrammatic LO (not NLO) result in a very similar way.
On the other hand, the perturbative result obtained from the exact expression of $\Gamma_\varphi$ does not even approach the diagrammatic result.

For the case of $n_\varphi=2$, the non-perturbative result approaches the diagrammatic prediction more slowly than for the $n_\varphi=1$ case.
For example, at $N=10$, the difference between the non-perturbative and diagrammatic results is about $2.5\%$ for $n_\varphi=1$, whereas it reaches roughly $20\%$ for $n_\varphi=2$.
The perturbative result based on the expanded expression of $\Gamma_\varphi$ approaches the diagrammatic result, but in a similar way to the $n_\varphi=1$ case rather than to the non-perturbative result for $n_\varphi=2$.
The perturbative result obtained with the exact expression for $\Gamma_\phi$ begins to deviate significantly from the diagrammatic prediction at a larger value of $N$ than for the $n_\varphi=1$ case.
This deviation becomes apparent around $N > 10$.

The results for $\mu A_\varphi = 0.3$ and $\mu A_\varphi = 0.1$ are shown in Figs.~\ref{fig:muphiew0p3} and \ref{fig:muphiew0p1}, respectively.
The qualitative behavior is the same as that seen for $\mu A_\varphi = 0.8$.
A noticeable difference appears in the case of $n_\varphi = 2$.
The agreement between the non-perturbative and diagrammatic results is achieved only for larger values of $N$ compared with the case of $\mu A_\varphi = 0.8$.
For $\mu A_\varphi = 0.8$, the ratio $\Gamma/\Gamma_{\rm dia}$ is already very close to unity at $N=100$.
In contrast, for $\mu A_\varphi = 0.3 \, (0.1)$, we find $N = 10^3 \, (10^4)$.
These values roughly coincide with those when the peak value of the phase-space density exceeds unity as seen in Fig.~\ref{fig:peak_value}, and thus scale as $1/(\mu A_\varphi)^2$ as expected from the growth rate of the parametric resonance [see the discussion around Eq.~\eqref{eq:k_res}].

On the other hand, the values for $n_\varphi = 1$ is $N = 100$ independently of $\mu A_\varphi$, though $1/(\mu A_\varphi)$ scaling is expected both from Fig.~\ref{fig:peak_value} and parametric resonance.
In other words, the non-perturbative result agrees with the diagrammatic result even before no peak experiences exponential growth yet.
Remembering that $n_\varphi = 1$ gives the leading-order contribution in perturbative theory, this indicates (and we also check) that the total (sum of all $n_\varphi$) decay rate agrees with the diagrammatic result for a large $N$ independently of $\mu A_\varphi$.
This is not very surprising, since the Feynman-diagrammatic approaches assume a large duration only in units of $1 / m_\varphi$ independently of $\mu A_\varphi$ [see discussion below \eqref{eq:GammaP}].
To examine this point further, we introduce a Rabi model in the next section.


\section{Rabi models}\label{sec:rabi}
To better understand our numerical results presented in the last section, we introduce a toy model, for which analytic results are available, in this section.
The time-dependent Hamiltonian is given by
\bal
\hat{H}(t) = \frac{1}{2} \omega \left(\hat{a} \hat{a}^\dagger + \hat{a}^\dagger \hat{a}\right) + \frac{1}{2} \gamma e^{i (m t - \phi)} \hat{a}^2 + \frac{1}{2} \gamma e^{- i (m t - \phi)} \hat{a}^{\dagger 2}  \,,
\eal
where $\omega$, $m$ and $\gamma > 0$ are real parameters whose units are the inverse of time (in natural units) and $\phi$ is a phase parameter.
The annihilation and creation operators satisfy
\bal
[\hat{a}, \hat{a}^\dagger] = 1 \,.
\eal
This is a Bosonic Rabi model, while a Fermionic (2-state) Rabi model can be found in standard textbooks (e.g., Ref.~\cite{Sakurai:2011zz}) and described in the appendix~\ref{sec:fermion}.

\subsection{Parametric-resonance approach}

One may see this analogy by noting that 
\bal
\hat{J}_+ = - i \frac{1}{2} \hat{a}^{\dagger 2} \,, \quad \hat{J}_- = - i \frac{1}{2} \hat{a}^2 \,, \quad \hat{J}_3 = \frac{1}{4} \left(\hat{a} \hat{a}^\dagger + \hat{a}^\dagger \hat{a}\right)
\eal
satisfy the complexified ($\hat{J}_\pm^\dagger = - \hat{J}_\mp$) Lie algebra of SU(2):
\bal
[\hat{J}_+, \hat{J}_-] = 2 \hat{J}_3 \,, \quad [\hat{J_3}, \hat{J}_\pm] = \pm \hat{J}_\pm \,,
\eal
(see, e.g., Ref.~\cite{Breczewski:2026utp} for a more general case with mode mixings).
Therefore, one can obtain the Bosonic result by changing $\gamma \to i \gamma$ and multiplying $- 1/2$ (to account for that the Fermionic model has two Fermionic degrees of freedom, while the Bosonic model has one Bosonic degree of freedom) in the Fermionic result. 
But, it would be still illustrative deriving the result as follows. 

The time-evolution (Unitary) operator is given by
\bal
&\hat{U}(t) = \hat{U}_{0}(t) \hat{U}_{I}(t) \,, \quad 
\hat{U}_{0}(t) = \exp\left[ - i \frac{m}{4} t \left(\hat{a} \hat{a}^\dagger + \hat{a}^\dagger \hat{a}\right) \right] \,, \notag \\
&\hat{U}_{I}(t) = \exp\left[ - i \left( \frac{\omega}{2} - \frac{m}{4} \right) t \left(\hat{a} \hat{a}^\dagger + \hat{a}^\dagger \hat{a}\right) - i \frac{1}{2} \gamma t e^{- i \phi} \hat{a}^2 - i \frac{1}{2} \gamma t e^{i \phi} \hat{a}^{\dagger 2} \right] \,.
\eal
One can check that it satisfies
\bal
i \frac{d}{dt} \hat{U}(t) = \hat{H} (t) \hat{U}(t) \,,
\eal
by using
\bal
\hat{U}_0(t) \hat{a} \hat{U}_0^\dagger(t) = e^{i m t / 2} \hat{a} \,, \quad \hat{U}_0(t) \hat{a}^\dagger \hat{U}^\dagger_0(t) = e^{- i m t / 2} \hat{a}^\dagger \,.
\eal
We consider the following time-dependent annihilation and creation operators:
\bal
\hat{b}(t) = \hat{U}_I(t) \hat{a} \hat{U}_I^\dagger(t) \,, \quad \hat{b}^\dagger(t) = \hat{U}_I(t) \hat{a}^\dagger \hat{U}_I^\dagger(t) \,,
\eal
which follow
\bal
i \frac{d}{dt} \hat{b}(t) = - \left( \omega - \frac{m}{2} \right) \hat{b}(t) - \gamma e^{i \phi} \hat{b}^\dagger(t) \,, \quad
i \frac{d}{dt} \hat{b}^\dagger(t) = \gamma e^{- i \phi} \hat{b}(t) + \left( \omega - \frac{m}{2} \right) \hat{b}^\dagger(t) \,.
\eal
One can check that the followings satisfy these equations:
\bal
& \hat{b}(t) = \left[ \cos(\Omega t) + i \frac{\sin(\Omega t)}{\Omega} \left( \omega - \frac{m}{2} \right) \right] \hat{a} + i \frac{\sin(\Omega t)}{\Omega} \gamma e^{i \phi}  \hat{a}^\dagger \,, \notag \\
& \hat{b}^\dagger(t) = \left[ \cos(\Omega t) - i \frac{\sin(\Omega t)}{\Omega} \left( \omega - \frac{m}{2} \right) \right] \hat{a}^\dagger - i \frac{\sin(\Omega t)}{\Omega} \gamma e^{- i \phi} \hat{a} \,,
\eal
which are even functions of $\Omega$ and thus depend only on (not on whether $\Omega$ is real or imaginary)
\bal
\Omega^2 = \left( \omega - \frac{m}{2} \right)^2 - \gamma^2 \,.
\eal
As a result, one obtains
\bal
&\hat{a}(t) = \hat{U}(t) \hat{a} \hat{U}^\dagger(t) = \mu(t) \hat{a} + \nu(t) \hat{a}^\dagger \,, \quad \hat{a}^\dagger(t) = \hat{U}(t) \hat{a}^\dagger \hat{U}^\dagger(t) = \nu^{*}(t) \hat{a} + \mu^{*}(t) \hat{a}^\dagger \notag \\
& \mu(t) = \left[ \cos(\Omega t) + i \frac{\sin(\Omega t)}{\Omega} \left( \omega - \frac{m}{2} \right) \right] e^{i m t / 2} \,, \quad \nu(t) = i \frac{\sin(\Omega t)}{\Omega} \gamma e^{i \phi} e^{- i m t / 2} \,.
\eal
One can check that
\bal
|\mu(t)|^2 - |\nu(t)|^2 = 1 \,,
\eal
and thus the inverse relation is given by
\bal
&\hat{a}_H(t) = \hat{U}^\dagger(t) \hat{a} \hat{U}(t) = \mu^*(t) \hat{a} - \nu(t) \hat{a}^\dagger \,, \quad \hat{a}^\dagger_H(t) = \hat{U}^\dagger(t) \hat{a}^\dagger \hat{U}(t) = - \nu^{*}(t) \hat{a} + \mu(t) \hat{a}^\dagger \,.
\eal
These relations are called the Bogoliubov transformation.

The state evolves from the vacuum,
\bal
|0\rangle (t) = \hat{U}(t) |0\rangle \,,
\eal
satisfies
\bal
\hat{a}(t) |0\rangle (t) = 0 \,.
\eal
One can check that it is given by the number operators,
\bal
| n \rangle = \frac{1}{\sqrt{n!}} \hat{a}^{\dagger n} | 0 \rangle \,,
\eal
as (up to the irrelevant overall phase $\alpha(t)$)
\bal
|0\rangle (t) = e^{i \alpha(t)} \frac{1}{\left(1 + |\nu(t)|^2 \right)^{1/4}} \sum_{m=0}^{\infty} \left( - \frac{\nu (t)}{\mu (t)} \right)^{m} \sqrt{\frac{(2m -1)!!}{(2m)!!}} |2m \rangle \,,
\eal
by using
\bal
\sum_{m=0}^{\infty} \frac{(2m -1)!!}{(2m)!!} x^{2m} = \frac{1}{\sqrt{1 - x^2}} \,.
\eal
Therefore, the vacuum-to-vacuum transition amplitude with the time interval $T$ is given by
\bal
|_{\rm out}\langle 0 | 0 \rangle_{\rm in}| = \frac{1}{\left(1 + |\nu(T)|^2 \right)^{1/4}} = \frac{1}{\left( 1 + \gamma^2 \sin^2 (\Omega T) / \Omega^2 \right)^{1/4}} \,.
\eal
Here note that the expectation number of produced particles is given by 
\bal
f(t) = \langle 0| \hat{a}^\dagger_H(t) \hat{a}_H(t) | 0 \rangle = |\nu(t)|^{2} \,,
\eal
and thus $|_{\rm out}\langle 0 | 0 \rangle_{\rm in}| = [1 + f(T)]^{-1/4}$.
We remark that for $|\omega - m/2| < \gamma$ (or, for imaginary $\Omega$), $f(t)$ grows exponentially.

We define the (possibly time-dependent) decay rate as
\bal
|_{\rm out}\langle 0 | 0 \rangle_{\rm in}| = \exp\left( - \Gamma(T) T / 2 \right)
\eal
and thus
\bal
\Gamma^P_\omega(T) = \frac{1}{2T} \ln \left[ 1 + f(T)\right] = \frac{1}{2T} \ln \left( 1 + \gamma^2 \sin^2 (\Omega T) / \Omega^2 \right) \,.
\eal


\subsection{Feynman-diagrammatic approach}

One can compute the vacuum-to-vacuum amplitude also in the Feynman-diagrammatic approach.
We split the Hamiltonian into the free and interaction parts:
\bal
\hat{H}(t) = \hat{H}_0 + \hat{V} (t) \,, \quad \hat{H}_0 = \frac{1}{2} \omega \left(\hat{a} \hat{a}^\dagger + \hat{a}^\dagger \hat{a}\right) \,, \notag \\
\hat{V} (t) = \frac{1}{2} \gamma e^{i (m t - \phi)} \hat{a}^2 + \frac{1}{2} \gamma e^{- i (m t - \phi)} \hat{a}^{\dagger 2}  \,.
\eal
Then $\hat{S}$ operator is in general given by
\bal
\hat{S} = T \left\{ \exp \left( - i \int dt \hat{V}_I (t) \right) \right\} \,, \hat{V}_I (t) = \exp \left(i \hat{H}_0 t \right) \hat{V}(t) \exp \left(-i \hat{H}_0 t \right) \,,
\eal
with $T$ denoting time-ordering.
In our case, there are outgoing-energy and incoming-energy vertices with energy $m$:
\bal
&\hat{V}_I (t) =  \hat{V}_- (t) + \hat{V}_+ (t) \,, \quad \hat{V}_- (t) = \frac{1}{2} \gamma e^{i (m t - \phi)} \hat{a}^2(t) \,, \quad \hat{V}_+ (t) = \frac{1}{2} \gamma e^{- i (m t - \phi)} \hat{a}^{\dagger 2}(t)\,, \notag \\
&\hat{a}(t) = \exp \left(i \hat{H}_0 t \right) \hat{a} \exp \left(-i \hat{H}_0 t \right) = e^{- i \omega t} \hat{a} \,, \quad \hat{a}^\dagger(t) = \exp \left(i \hat{H}_0 t \right) \hat{a}^\dagger \exp \left(-i \hat{H}_0 t \right) = e^{i \omega t} \hat{a}^\dagger \,.
\eal

The exponent of vacuum-to-vacuum amplitude is computed as
\bal
_{\rm out}\langle 0 | 0 \rangle_{\rm in} = \langle 0 | \hat{S} | 0 \rangle \,.
\eal
Through the Wick theorem, time-ordered products of annihilation and creation operators are given by normal-ordered products + all possible contractions.
The contraction gives us a propagator. In our case,
\bal
&\langle 0 | T\left\{ \hat{a}(t_1) \hat{a}^\dagger(t_2) \right\} | 0 \rangle = \langle 0 | T\left\{ \hat{a}^\dagger(t_2) \hat{a}(t_1) \right\} | 0 \rangle \notag \\
&= \theta(t_1 - t_2) e^{- i \omega (t_1 - t_2)}  = \int \frac{d \omega'}{2 \pi} \frac{i}{\omega' - \omega + i \epsilon} e^{- i \omega' (t_1 - t_2)} \,.
\eal
With this Feynman rule (vertex is the same as usual quantum field theory), one can compute the decay rate by the imaginary part of the sum of the connected diagrams with all possible insertions of vertices.

\begin{figure}
    \centering
    \includegraphics[width=0.46\linewidth]{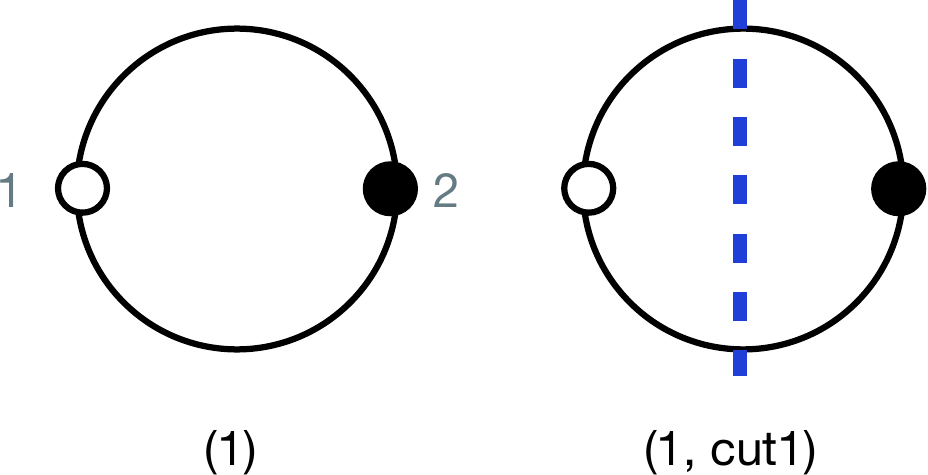}
    \hspace{7mm}
    \includegraphics[width=0.46\linewidth]{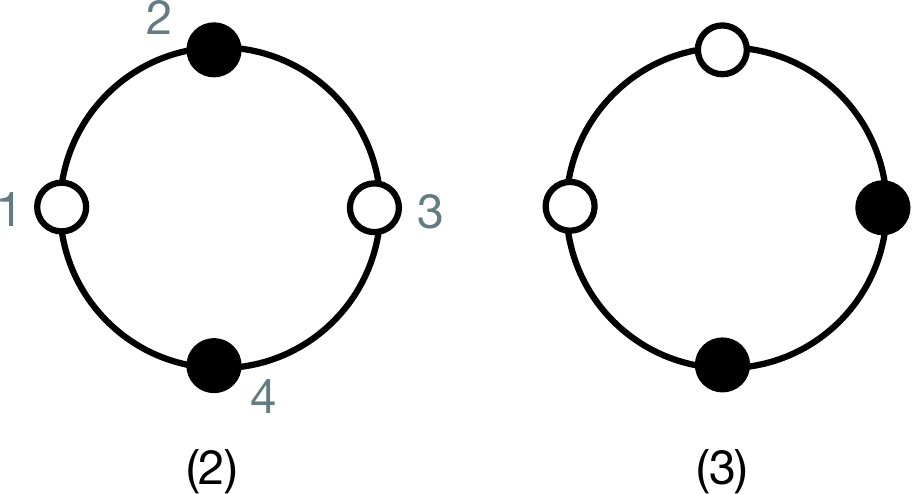}
    \caption{Feynman diagrams for the Rabi model. These diagrams are quoted from Fig.~3 and Fig.~4 in Ref.~\cite{Kamada:2025evo}. In the Rabi model, the black and white circles should be paired side by side. Hence, in this model, the diagram (3) does not exist.}
    \label{fig:diagrams}
\end{figure}

Noting that outgoing-energy and incoming energy vertices have only $\hat{a}$ and $\hat{a}^\dagger$, respectively, one finds that they should be inserted one after the other. We denote the number of pairs by $p$.
The amplitude is given by
\bal
i {\cal M}_{p} = \frac{1}{2p}\gamma^{2p} \int \frac{d \omega'}{2 \pi} (\omega' - \omega + i \epsilon)^{-p} (m - \omega' - \omega + i \epsilon)^{-p} \,,
\eal
where the prefactor $1/2p$ is a symmetric factor to avoid double-counting.
Note that the adiabatic parameter does not need to be the same for all the propagators, though we denote it in the same symbol for brevity.
Therefore, for example, $(\omega' - \omega + i \epsilon)^{-p}$ is not a pole of order $p$ (if so, the amplitude would vanish for $p > 1$ according to Cauchy integral theorem).
For example, the diagrams for $p=1$ and $p=2$ are shown in Fig.~\ref{fig:diagrams}.  

To avoid this notational problem and also simplify computation, we employ the ``pinch'' technique.
With it, the amplitude is rewritten as
\bal
i {\cal M}_p = \frac{1}{2p}\gamma^{2p} \frac{1}{[(p-1)!]^2} \lim_{\omega_1\to \omega} \lim_{\omega_2\to \omega} \frac{\partial^{p-1}}{\partial \omega_1^{p-1}} \frac{\partial^{p-1}}{\partial \omega_2^{p-1}} \int \frac{d \omega'}{2 \pi} (\omega' - \omega_1 + i \epsilon)^{-1} (m - \omega' - \omega_2 + i \epsilon)^{-1} \,.
\eal
One can perform the integral
\bal
\int \frac{d \omega'}{2 \pi} (\omega' - \omega_1 + i \epsilon)^{-1} (m - \omega' - \omega_2 + i \epsilon)^{-1} = - \pi \delta(m - \omega_1 - \omega_2) - i PV (m - \omega_1 - \omega_2)^{-1} \,,
\eal
with $PV$ denoting the principal value.
As a result, the imaginary part of the amplitude is given by
\bal
{\rm Im} {\cal M}_p = \pi \frac{1}{2p}\gamma^{2p} \frac{1}{[(p-1)!]^2} \lim_{\omega_1\to \omega} \lim_{\omega_2\to \omega} \frac{\partial^{p-1}}{\partial \omega_1^{p-1}} \frac{\partial^{p-1}}{\partial \omega_2^{p-1}} \delta(m - \omega_1 - \omega_2) \,.
\eal

Though this computation is good enough for our purpose, it is still illuminating obtaining the result by applying cutting rules.
The discontinuity of the Feynman amplitude is computed by the  replacement: for example,
\bal
(\omega' - \omega + i \epsilon)^{-1} \to (- 2 \pi i) \delta(\omega' - \omega) \,,
\eal
for cut propagators.
By applying it to the pinched amplitude, one obtains
\bal
{\rm Disc.} (i {\cal M}_p) = -   2 \pi \frac{1}{2p}\gamma^{2p} \frac{1}{[(p-1)!]^2} \lim_{\omega_1\to \omega} \lim_{\omega_2\to \omega} \frac{\partial^{p-1}}{\partial \omega_1^{p-1}} \frac{\partial^{p-1}}{\partial \omega_2^{p-1}} \delta(m - \omega_1 - \omega_2) \,.
\eal
We confirm the standard relation: ${\rm Disc.} (i {\cal M}) = - 2 {\rm Im}{\cal M}$. 

One obtains the decay rate
\bal
\Gamma^F_\omega = 2 {\rm Im}{\cal M} = \sum_{p=1}^\infty 2 \pi \frac{1}{2p}\gamma^{2p} \frac{1}{[(p-1)!]^2} \lim_{\omega_1\to \omega} \lim_{\omega_2\to \omega} \frac{\partial^{p-1}}{\partial \omega_1^{p-1}} \frac{\partial^{p-1}}{\partial \omega_2^{p-1}} \delta(m - \omega_1 - \omega_2) \,.
\eal
The first equality follows from that the delta function for the energy conservation, which is multiplied by the amplitude, turns into $T$, since a bubble diagram trivially satisfies the energy conservation.

\subsection{Perturbative comparison of two approaches}
Let us compare two results order by order in terms of $\gamma^2$.
In the Feynman-diagrammatic approach, $\gamma^{2p}$-order contribution can be easily find as
\bal
\Gamma^F_{p, w} = 2 \pi \frac{1}{2p}\gamma^{2p} \frac{1}{[(p-1)!]^2} \lim_{w_1\to w} \lim_{w_2\to w} \frac{\partial^{p-1}}{\partial w_1^{p-1}} \frac{\partial^{p-1}}{\partial w_2^{p-1}} \delta(w_1 + w_2) \,,
\eal
where $w = \omega - m/2$.
Since it is not easy to compare $\delta$ function (and its derivative) directly, we consider a test function $f(w)$:
\bal
\int dw f(w) \Gamma^F_{p, w} &= 2 \pi \frac{1}{2p}\gamma^{2p} \frac{1}{[(p-1)!]^2} \int dw \lim_{w_1\to w} \lim_{w_2 \to w} f\left( \frac{w_{1} + w_{2}}{2}\right) \frac{\partial^{p-1}}{\partial w_1^{p-1}} \frac{\partial^{p-1}}{\partial w_2^{p-1}} \delta(w_1 + w_2) \notag \\
&= 2 \pi \frac{1}{2p}\gamma^{2p} \frac{1}{[(p-1)!]^2} \frac{1}{2^{2p-2}}\int dw \left[ \frac{d^{2p-2}}{d w^{2p-2}} f(w) \right] \delta(2 w) \notag \\
&= 2 \pi \gamma^{2p} \frac{1}{p!(p-1)!} \frac{1}{2^{2p}} \frac{d^{2p-2}}{d w^{2p-2}} f(w=0) \,.
\label{eq:integral}
\eal

In the parametric-resonance approach, the perturbative computation is performed in the following way (though we discuss a general case first, it may be easier to see low-order results below at the same time).
The $\gamma^{2p}$-order contribution is decomposed into
\bal
\Gamma^P_{p, w} = \Gamma^{P, +}_{p, w} + \Gamma^{P, -}_{p, w} \,,
\eal
where $\Gamma^{P, +}_{p, w}$ has only negative frequency contributions:
\bal
\Gamma^{P, +}_{p, w} = \sum_{n=0}^{p} W_{p, n}(w) e^{2 i n w T} \,,
\eal
with $W_{p, n}(w)$ is a rational function.
$\Gamma^{P, -}_{p, w} = \Gamma^{P, +}_{p, -w}$ has only positive frequency contributions.
(Remember that $\Gamma^P_{p, w}$ is an even function of $w$.)
Suppose $f(w)$ is a ``good'' function to justify the following derivation (we will discuss this point in the end of this section).
Since $\Gamma^P_{p, w}$ (but not $\Gamma^{P, +}_{p, w}$ nor $\Gamma^{P, -}_{p, w}$) is regular at the origin $w=0$, one can modify the integration contour so that it circumvents the origin counter-clockwise.
Then, for $f(w) \Gamma^{P, +}_{p, w}$, one can close the contour in the upper-half plane and then apply the residue theorem at the origin.
Meanwhile, for $f(w) \Gamma^{P, -}_{p, w}$, one can close the contour in the lower-half plane, to obtain 0.
In summary, the integral is given by
\bal
\int dw f(w) \Gamma^P_{p, w} = 2 \pi i {\rm Res} \left[f(w) \Gamma^{P, +}_{p, w}, 0\right] \,,
\label{eq:Res}
\eal
with ${\rm Res} \left[F(z), z_0\right]$ denoting the residue of $F(z)$ at $z = z_0$, which is zero if there is no single pole at $z = z_0$.

For this residue to agree with the Feynman-diagrammatic result \eqref{eq:integral}, one expects the following Laurent expansion of $\Gamma^{P, +}_{p, w}$ around $w = 0$:
\bal
\Gamma^{P, +}_{p, w} = - i \frac{(2p-3)!!}{2^{p+1}p!} \frac{1}{w^{2p-1}}  \gamma^{2p}+ {\cal O}(w^0)\gamma^{2p}.
\label{eq:pole}
\eal
One can find the leading 3 orders as
\bal
& \Gamma^P_{1, w} = \frac{1}{2 w^2 T} \sin^2(wT) \gamma^2 \,, \notag \\
& \Gamma^P_{2, w} = \frac{1}{4 w^4 T} \left( - 2 w T \cos(wT) \sin(wT) + 2 \sin^2(wT) - \sin^4(wT) \right) \gamma^4 \,, \notag \\
& \Gamma^P_{3, w} = \frac{1}{24 w^6 T} \left[ 3 (w T)^2 \cos^2(wT) - 15 wT \cos(wT) \sin(wT) + [12 - 3 (wT)^2] \sin^2(wT) \right. \notag \\
& \qquad \qquad \qquad \quad \left. + 12 wT \cos(wT) \sin^3(wT) - 12 \sin^4(wT) + 4 \sin^6(wT)\right] \gamma^6  \,,
\eal
to confirm
\bal
& \Gamma^{P,+}_{1, w} = \frac{1}{2 w^2 T} \frac{e^{2iwT} - 1}{(2 i)^2} \gamma^2 = - i \frac{1}{4} \frac{1}{w} \gamma^2 + \frac{1}{4} \gamma^2 T + \dots \,, \notag \\
& \Gamma^{P,+}_{2, w} = \frac{1}{4 w^4 T} \left( - 2 w T \frac{e^{2iwT}}{2 (2 i)} + 2 \frac{e^{2iwT} - 1}{(2 i)^2} - \frac{e^{4iwT} - 4 e^{2iwT} +3}{(2 i)^4} \right) \gamma^4 \notag \\
& \qquad \, = - i \frac{1}{16} \frac{1}{w^3} \gamma^4 - \frac{1}{24} \gamma^4 T^{3} + \dots \,, \notag \\
& \Gamma^{P,+}_{3, w} = \frac{1}{24 w^6 T} \left( 3 (w T)^2 \frac{e^{2iwT} + 1}{2^2} - 15 wT \frac{e^{2iwT}}{2 (2 i)} + [12 - 3 (wT)^2] \frac{e^{2iwT} - 1}{(2 i)^2} \right. \notag \\
& \qquad \qquad \qquad \quad \left. + 12 wT \frac{e^{4iwT} -2 e^{2iwT}}{2 (2 i)^3} - 12 \frac{e^{4iwT} - 4 e^{2iwT} +3}{(2 i)^4} + 4 \frac{e^{6iwT} - 6 e^{4iwT} + 15 e^{2iwT} - 10}{(2 i)^6} \right) \gamma^6  \notag \\
& \qquad \, = - i \frac{1}{32} \frac{1}{w^5} \gamma^6 + \frac{1}{90} \gamma^6 T^5 + \dots
\eal
In the last equalities, we Laurent expand the expressions around $w = 0$.

We have seen that $\int d w f(w) \Gamma^P_\omega = \int d w f(w) \Gamma^F_\omega$ at all orders in perturbation theory.
But remember that in the derivation of \eqref{eq:Res}, we assume that  $f(w)$ is a ``good'' function.
Actually, $f(w)$ needs to be analytic and regular everywhere in a complex $w$ plane.
From Liouville theorem, such a function is limited to be only a constant. (Actually we consider a constant case later.)
To loosen this limitation, we need to take a large $T$ limit (note that our derivation of \eqref{eq:Res} holds for any $T$).
We change the integration variable from $w$ to $y = w T$.
This leads to $f(w)=f(y/T)$ while $\Gamma^{P}_{p, w} = \gamma^{2p} T^{2p-1} G^{P}_{p} (y)$, where $G^{P}_{p} (y)$ is a function only of $y$.
At the large $T$ limit, we expect that $f(w)$ is well approximated by the Taylor series up to $(y/T)^{2p-2}$ (remember that an additional $T$ appears from the Jacobian).
This leads to the same result as \eqref{eq:Res} (especially, one can close the contour with a polynomial up to $w^{2p-2}$) and thus only requires that $f(w)$ is regular at the origin.
We admit that this argument is not mathematically rigorous, but a similar argument is commonly found in physics.
For example, a similar argument leads to $\sin^2(w T) / w^2 T \to \pi \delta(w)$ at a large $T$, which we used before. (Actually our computation with $p=1$ concerns the same asymptotic form).

\subsection{Non-perturbative comparison and implications}

Though $\Gamma^P_\omega = \Gamma^F_\omega$ seems to hold at a large $T$ perturbatively, we still wonder if it holds non-perturbatively.
We consider the integral of \eqref{eq:integral} again.
In the parametric-resonance approach,
\bal
\int dw f(w) \Gamma^P_w = \frac{1}{2 T} \int dw f(w) \ln \left[ 1 + \gamma^2 \sin^2 (\sqrt{w^2 - \gamma^2} T) / (w^2 - \gamma^2) \right] \,.
\eal
This would require the numerical computation for a general $T$, but actually, one can find an asymptotic form at a large $T$, noting that $|w| < \gamma$ (or, imaginary argument of a sign function) is a growing mode:
\bal
\sin^2 (\sqrt{w^2 - \gamma^2} T) / (w^2 - \gamma^2) = \sinh^2 (\sqrt{\gamma^2 - w^2} T) / (\gamma^2 - w^2) \approx e^{2 \sqrt{\gamma^2 - w^2} T} / 4 (\gamma^2 - w^2) \,,
\eal
and thus
\bal
\int dw f(w) \Gamma^P_w \approx \int_{- \gamma}^\gamma dw f(w) \sqrt{\gamma^2 - w^2} = 2 \pi \sum_{p=1}^{\infty} \gamma^{2p} \frac{1}{p!(p-1)!} \frac{1}{2^{2p}} \frac{d^{2p-2}}{d w^{2p-2}} f(w=0) \,.
\eal
In the second equality, we have Taylor-expanded $f(w)$ and used
\bal
\int_{0}^{\frac{\pi}{2}} dx \cos^{2p-2}x \sin^2x = \frac{\pi}{4} \frac{(2p-3)!!}{2^{p-1} p!} \,.
\eal
One can see that the parametric-resonance and Feynman-diagrammatic results coincide at a large $T$.
For a general $T$, one can rewrite the integral with $w = \gamma x $ and $\tau = \gamma T$ in the parametric-resonance approach:
\bal
\int dw f(w) \Gamma^P_w = \frac{\gamma^2}{2\tau} \int dx f(\gamma x) \ln \left[ 1 + \sin^2 (\sqrt{x^{2}-1} \tau) / (x^2 - 1) \right] \,.
\eal

\begin{figure}
    \centering
     \includegraphics[width=0.48\linewidth]{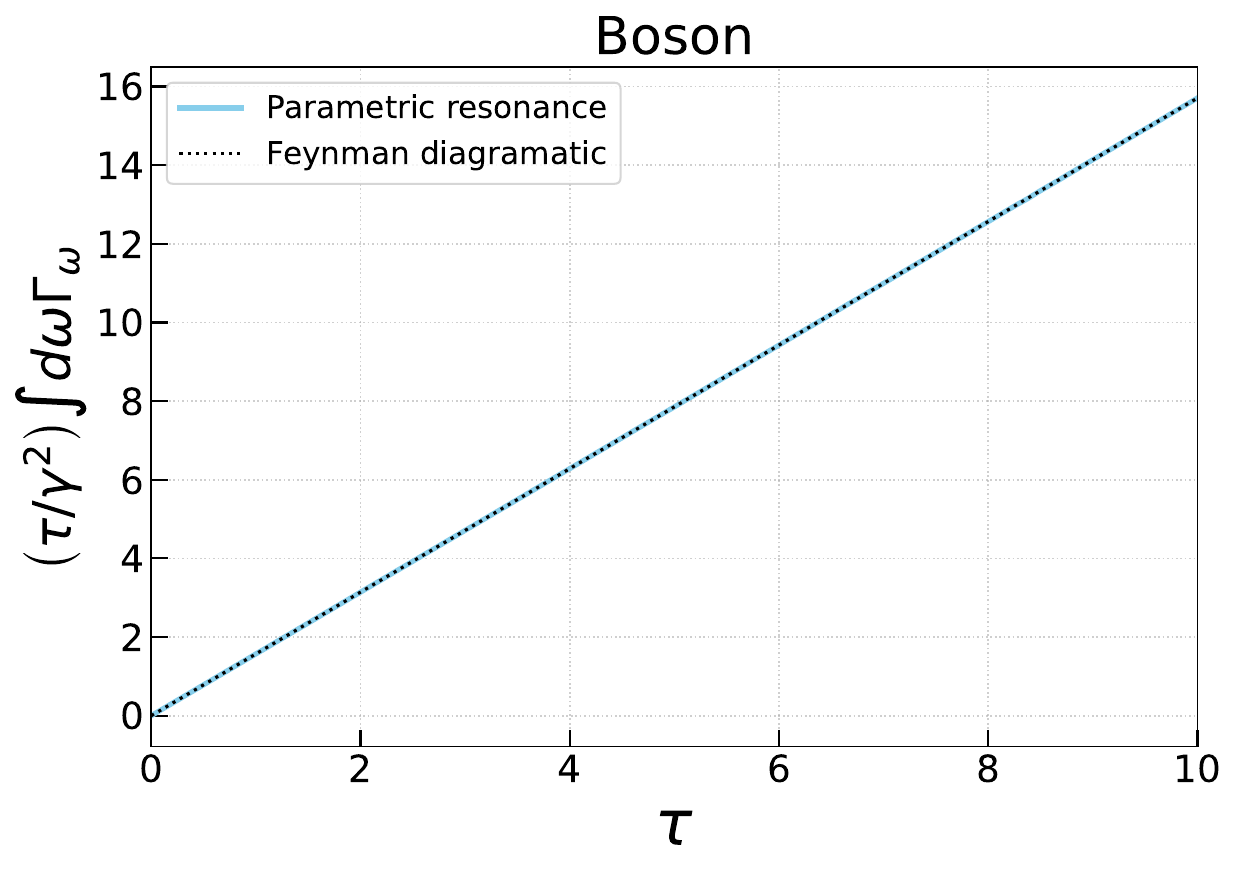}
     \includegraphics[width=0.48\linewidth]{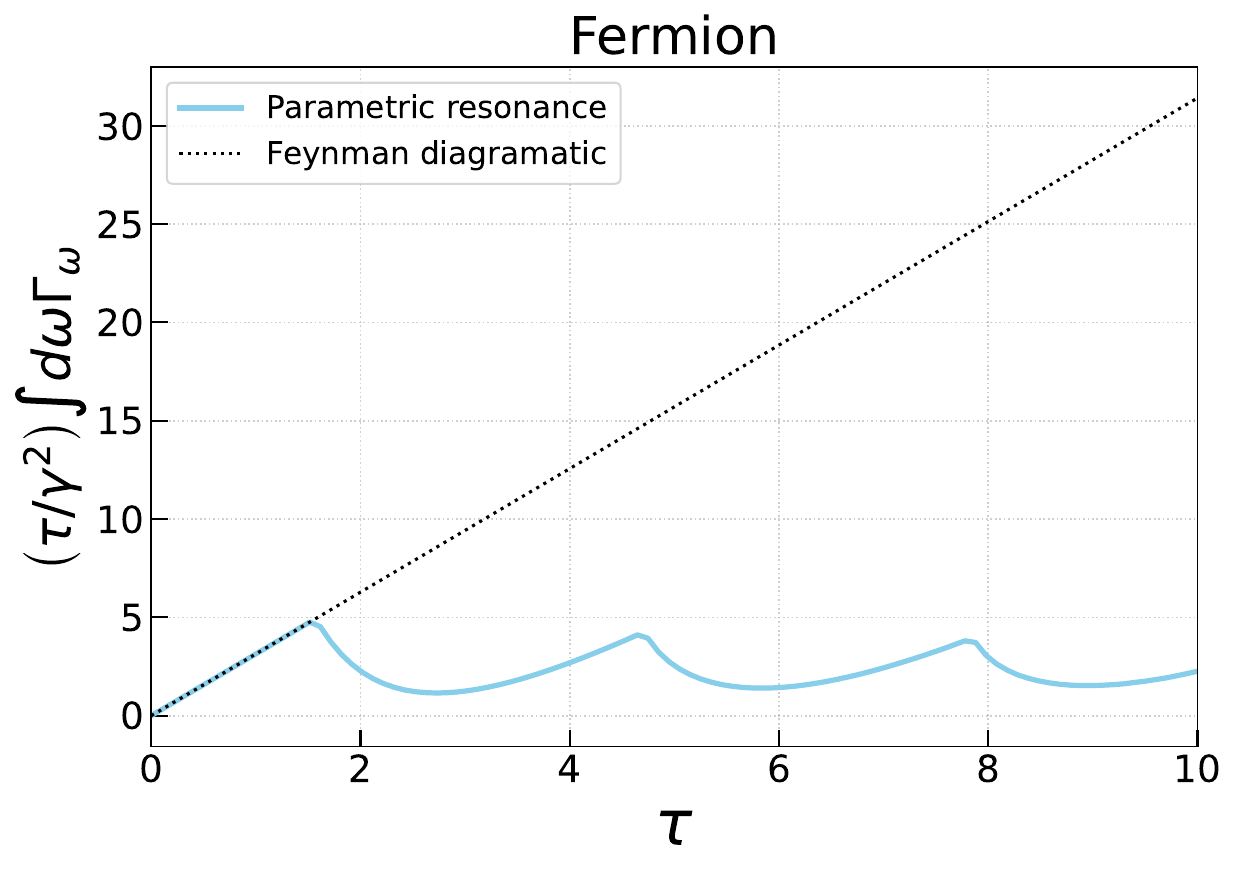}
    \caption{Decay rate as a function of $\tau$ in the Bosonic (left) and Fermionic (right) Rabi models.}
    \label{fig:Rabi1}
\end{figure}

First, we consider $f(w)=1$.
The numerical-integration result is presented in Fig.~\ref{fig:Rabi1} (left) and compared with the Feynman-diagrammatic approach (actually, one can perform the integration analytically; see appendix~\ref{sec:integral}):
\bal
\int dw \Gamma^F_w = \frac{\pi}{2} \gamma^2 \,.
\eal
The total decay rate in the Bosonic Rabi model agrees with each other for any time duration in the parametric-resonance and Feynman-diagrammatic approaches.
This agreement is expected from the discussion in the end of the last section.
This is the case where the derivation of \eqref{eq:Res} holds for any $T$.

\begin{figure}
    \centering
    \includegraphics[width=0.48\linewidth]{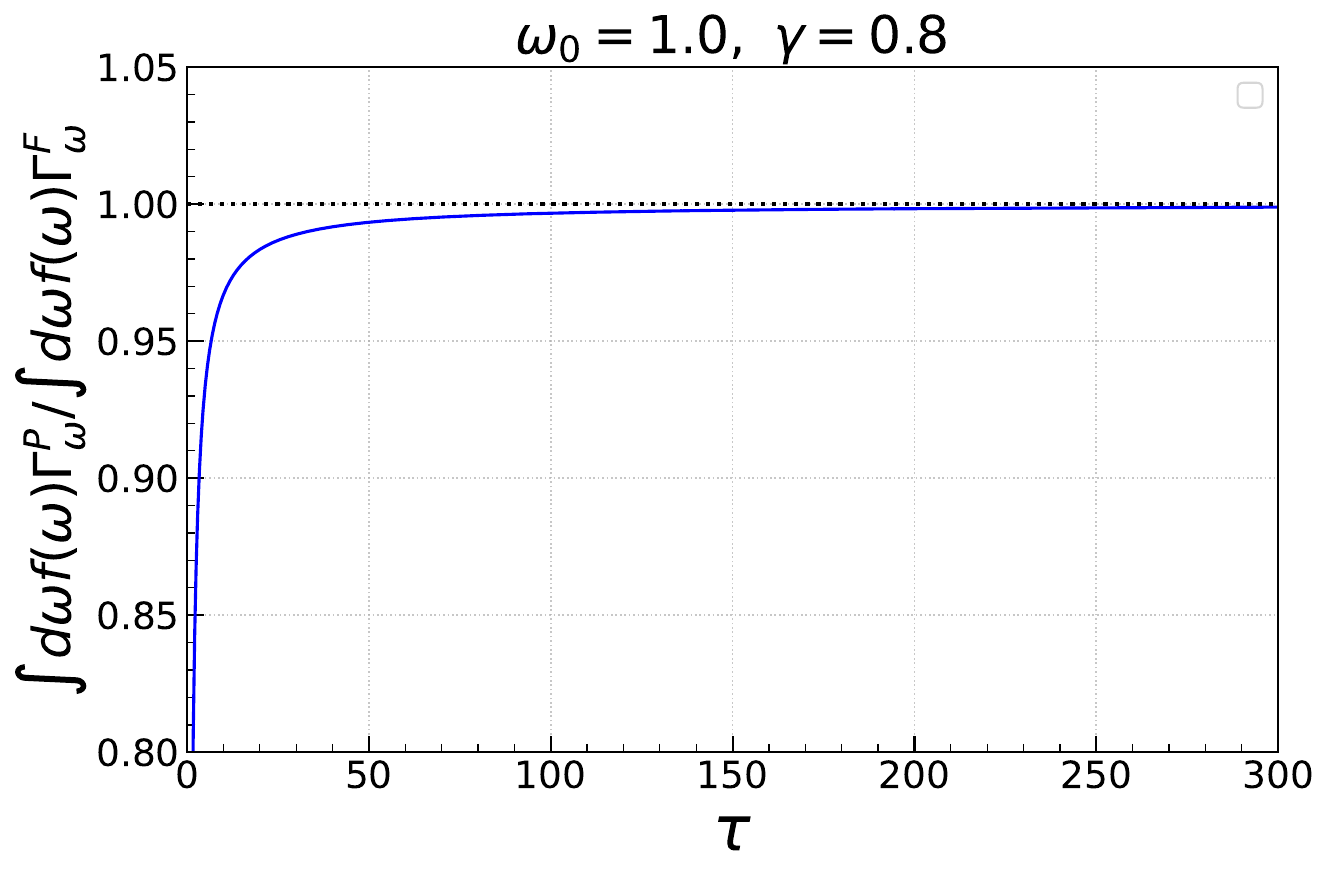}
   \includegraphics[width=0.48\linewidth]{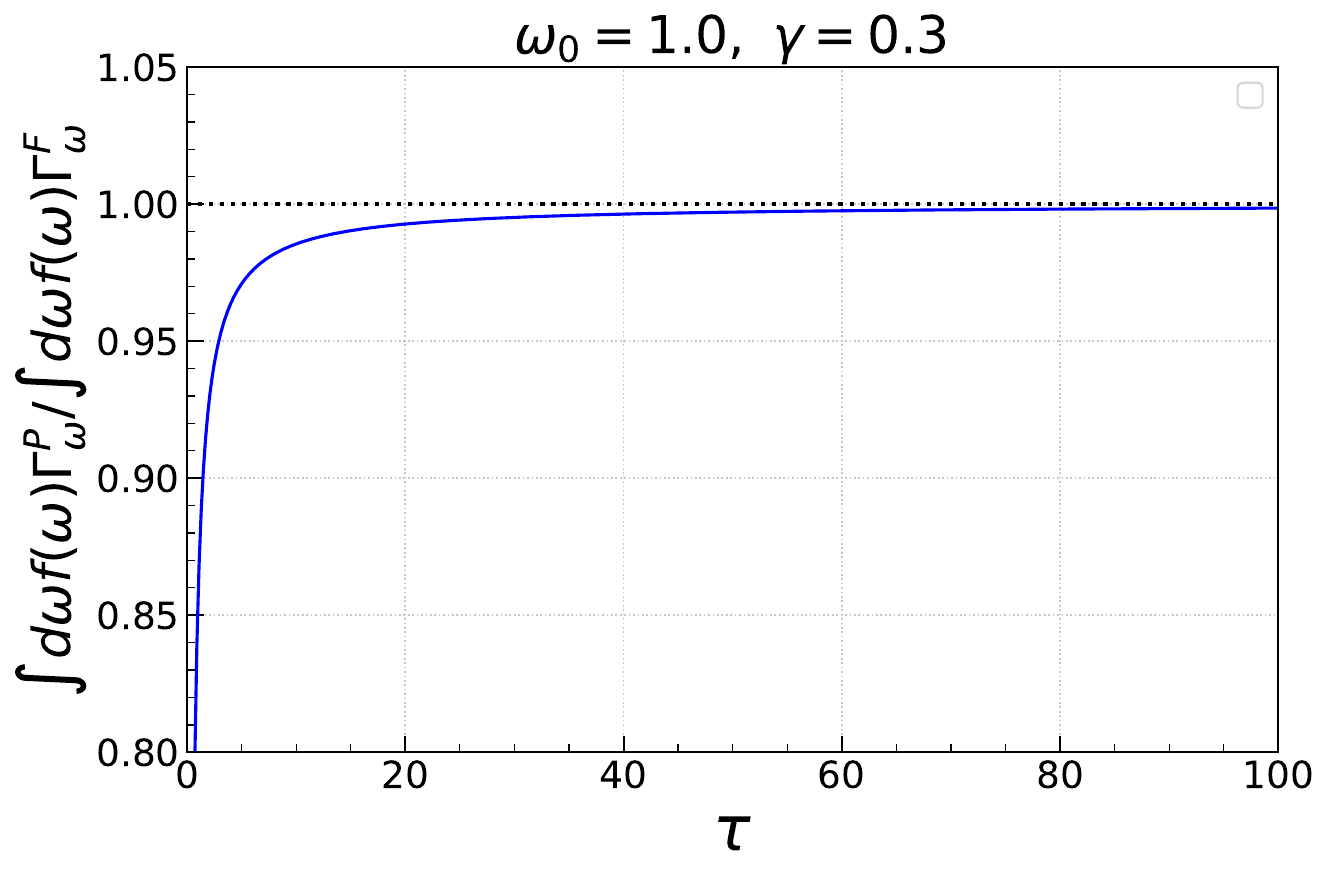}
    \includegraphics[width=0.48\linewidth]{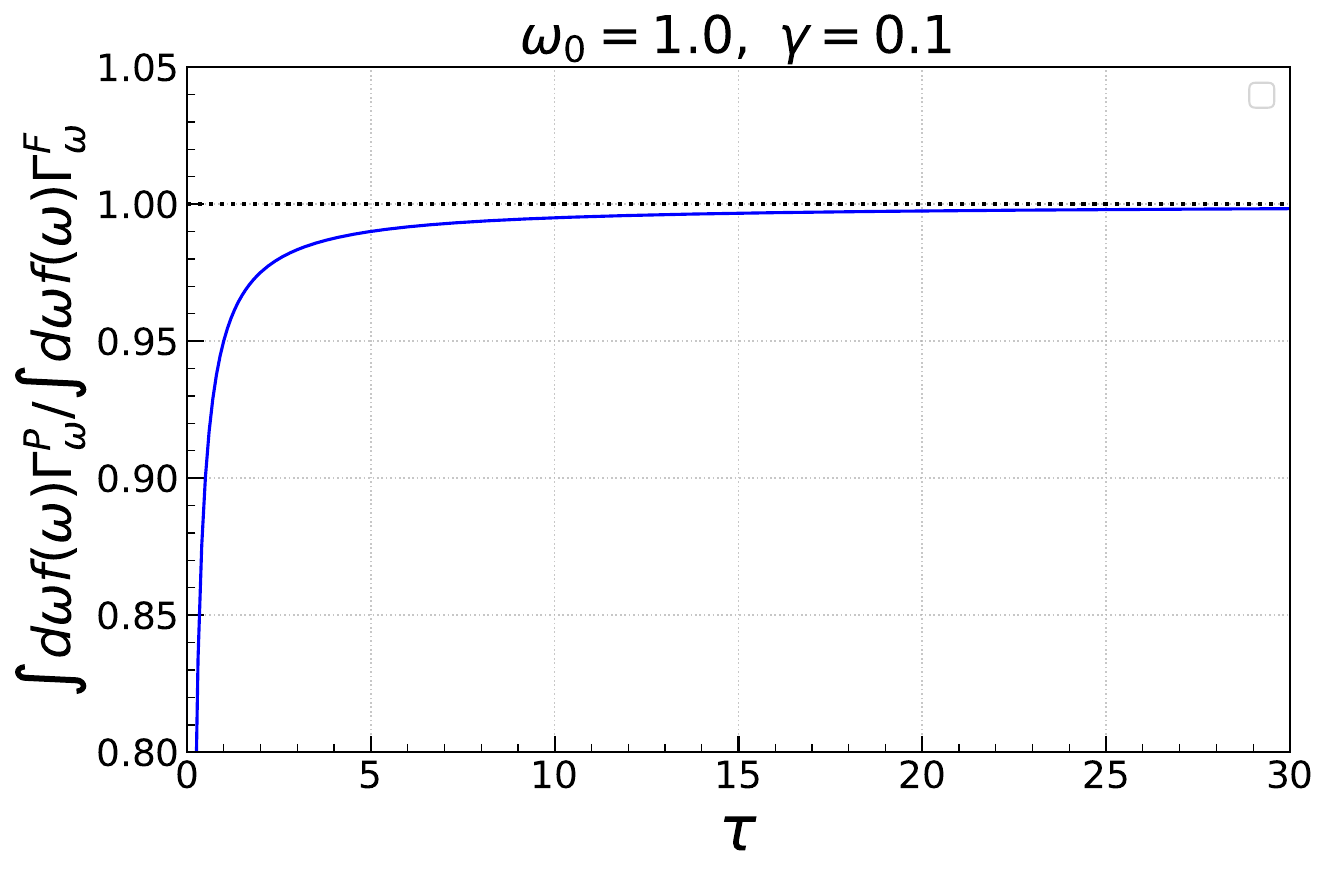}
    \caption{Ratio of $\int dw f(w)\Gamma_w$ obtained by the parametric resonance approach to that obtained by the Feynman diagrammatic approach as a function of $\tau$, with $f(w)=w_0^2 / (w_0^2 + w^2)$ in units of $w_0 = 1$.}
    \label{fig:Rabi2}
\end{figure}

Second, we consider $f(w)=w_0^2 / (w_0^2 + w^2)$ (we take the units of $w_0 = 1$ in numerical computations).
The numerical-integration result is presented in Fig.~\ref{fig:Rabi2} and compared with the Feynman-diagrammatic approach:
\bal
\int dw \frac{w_0^2}{w_0^2 + w^2} \Gamma^F_w = \pi \gamma^2 \frac{1}{\sqrt{1 + (\gamma/w_0)^2} + 1} \,.
\eal
We see a disagreement between parametric-resonance and Feynman-diagrammatic results for a small $\tau$, though the former approaches to the latter at a large $\tau$.
The dimensionless time duration, at which the agreement is achieved, depends on the coupling, $\tau \propto \gamma$, and thus corresponding $T$ does not depend on $\gamma$.
This agrees with the results in quantum field theory [see Figs.~\ref{fig:muphiew0p8}-\ref{fig:muphiew0p1}].

\begin{figure}
    \centering
    \includegraphics[width=1.\linewidth]{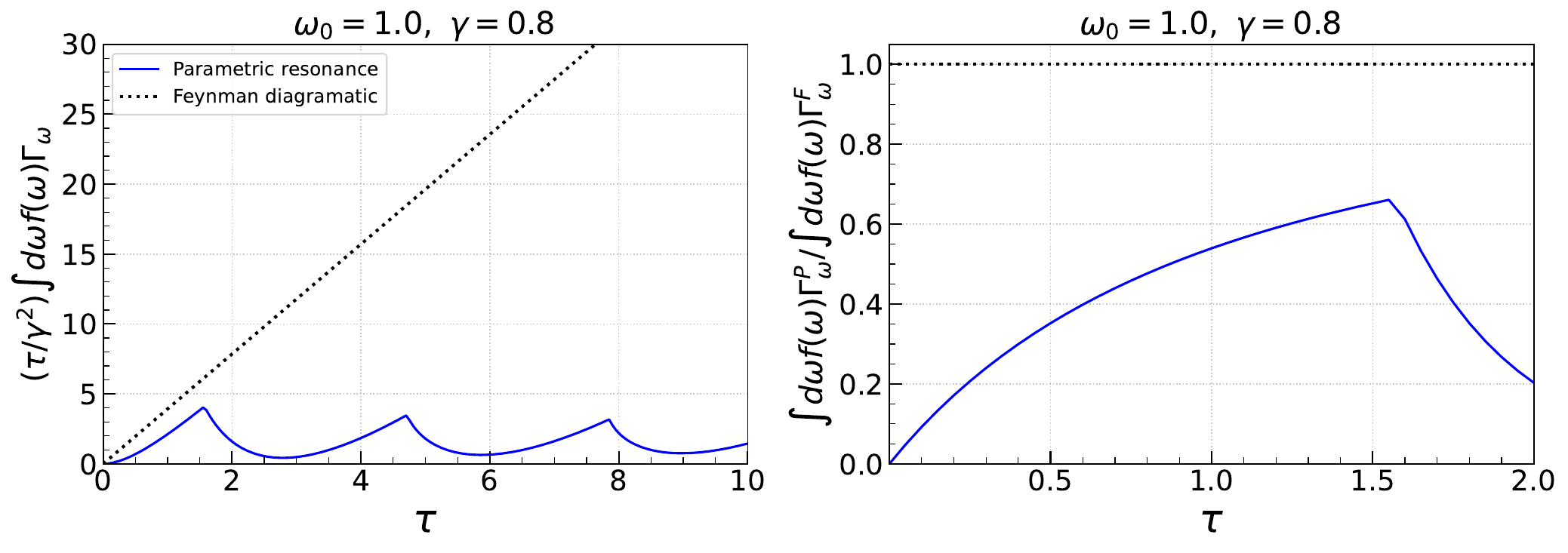}
   \includegraphics[width=1.\linewidth]{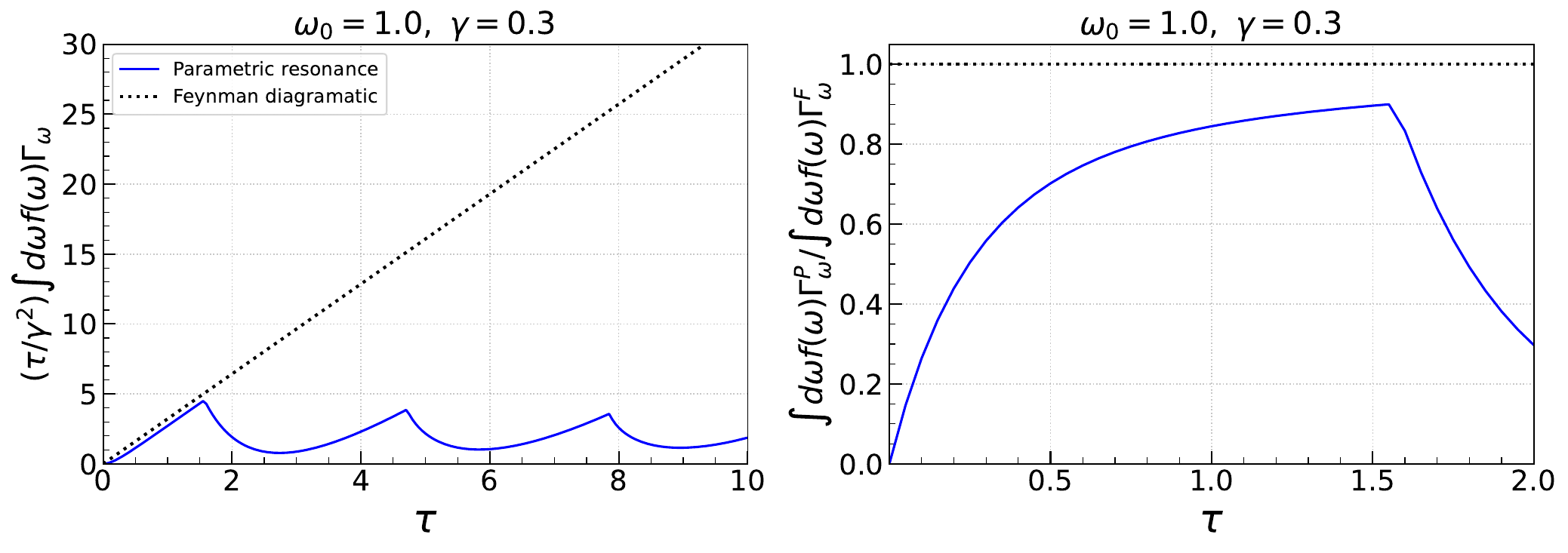}
    \includegraphics[width=1.\linewidth]{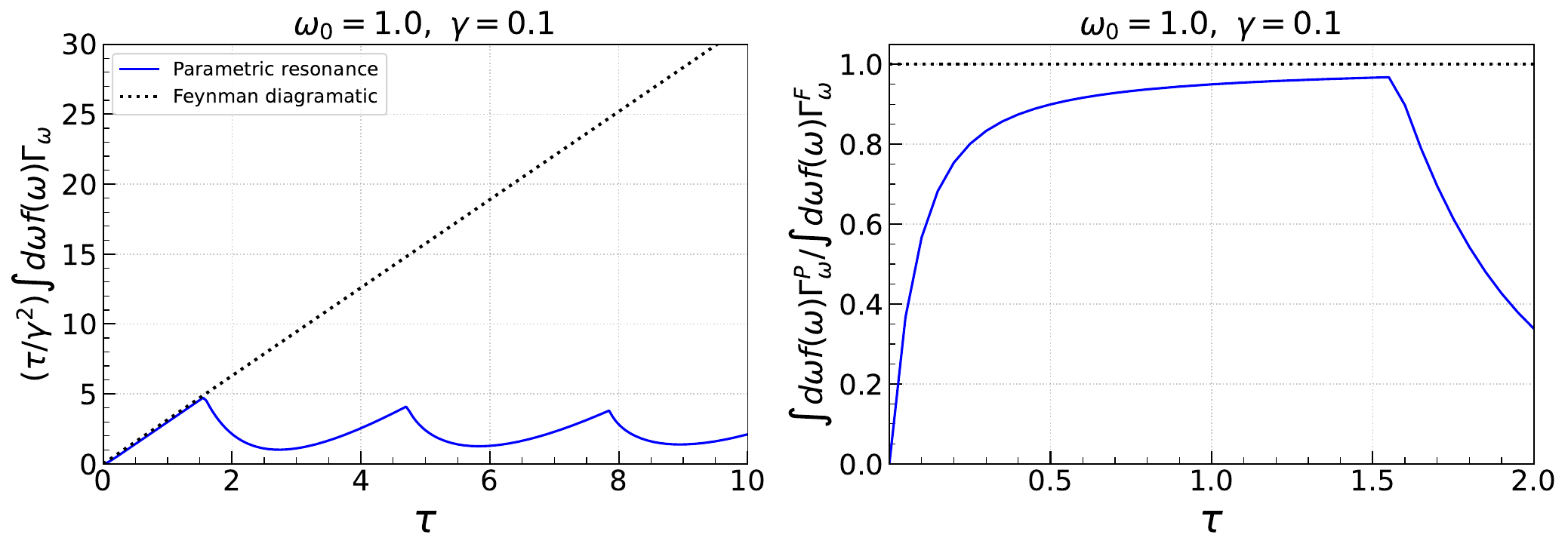}
    \caption{Numerical results of $ (\tau/\gamma^2)\int dw f(w)\Gamma_w$ as a function of $\tau$, with $f(w)=w_0^2 / (w_0^2 + w^2)$ in units of $w_0 = 1$.}
    \label{fig:Rabi_f}
\end{figure}

The result for $f(w)=1$ in the Fermionic model (see appendix~\ref{sec:fermion}) is plotted in Fig.~\ref{fig:Rabi1} (right).
The parametric-resonance and Feynman-diagrammatic results agree with each other for a small time duration, but not for a large time $\tau > \pi/2$.
This disagreement at a large time is surprising in two folds.
First, this is the case where the derivation of \eqref{eq:Res} holds for any $T$ as in the Bosonic case.
Second, the parametric-resonance and Feynman-diagrammatic results agree at a large time at all orders of perturbation theory.
On the other hand, physically this disagreement at a large time would be because of Pauli blocking.
As the produced number of Fermions increases, further decay is suppressed until the produced Fermions disappear via a production of the mother particle ($a^2$ term in the Hamiltonian).

The result for $f(w)=w_0^2 / (w_0^2 + w^2)$ in the Fermionic model is plotted in Fig.~\ref{fig:Rabi_f}.
Again the parametric-resonance and Feynman-diagrammatic results do not agree with each other for a large time $\tau > \pi/2$.
They do not agree for a small time, but the parametric-resonance approaches to the Feynman-diagrammatic result.
Though the approach ends abruptly at $\tau = \pi/2$, till then it seems that the time duration, at which the agreement is achieved, is proportional to $\gamma$, as in the Bosonic model [see Fig.~\ref{fig:Rabi2}].

\section{Conclusions}\label{sec:conclusions}

Motivated by Ref.~\cite{Kamada:2025evo}, we have studied the following two aspects of decay of scalar condensation.
First, we have studied the time evolution of decay rate for a Bosonic daughter particle based on the parametric-resonance approach.
More specifically, we have followed the time evolution of mode functions both analytically in perturbation theory and numerically.
We have computed the time evolution of the phase-space distribution based on the obtained mode functions.
The phase-space distribution exhibits peaks labeled by $n_\varphi$, as expected from the narrow resonance regime of the Mathieu equation.
The decay rate for each $n_\varphi$ approaches with time to the constant value obtained in the Feynman-diagrammatic approach.
The duration required to see their agreement scales with a coupling as expected from the growth rate of the parametric resonance for $n_\varphi = 2$.

On the other hand, it is independent of the coupling for $n_\varphi = 1$ unlike the expectation.
This difference would infer that the total decay rate agrees with the Feynman-diagrammatic approach after tens of oscillations independently of the coupling.
This would be well encapsulated by a Bosonic Rabi model in quantum mechanics.
We have derived the decay rate both non-perturbatively and Feynman-diagrammatically (perturbatively).
We have confirmed that the decay rate agrees with the Feynman-diagrammatic approach actually for all the time.

Second, we have studied the time evolution of decay rate for a Fermionic daughter particle.
Though we leave a study in quantum field theory like Ref.~\cite{Kamada:2025evo} and section~\ref{sec:calc} for a future work, we have taken a glimpse from a Fermionic Rabi model in quantum mechanics.
There is no qualitative change from the Bosonic case in perturbation theory as expected.
On the other hand, the non-perturbative decay rate does not agree with the Feynman-diagrammatic result for a large time duration.
This would be because of Pauli blocking and motivate a further study in quantum field theory.

\appendix

\section{Fermionic (2-state) Rabi model} \label{sec:fermion}

The time-dependent Hamiltonian of the Fermionic Rabi model is given by
\bal
\hat{H}(t) = \omega \left(\hat{a}^\dagger \hat{a} - \hat{c} \hat{c}^\dagger\right) + \gamma e^{i (m t - \phi)} \hat{c} \hat{a} + \gamma e^{- i (m t - \phi)} \hat{a}^\dagger \hat{c}^\dagger \,,
\eal
where $\omega$, $m$ and $\gamma > 0$ are real parameters whose units are the inverse of time (in natural units) and $\phi$ is a phase parameter.
The Fermionic annihilation and creation operators satisfy
\bal
\{\hat{a}, \hat{a}^\dagger\} = \{\hat{c}, \hat{c}^\dagger\} = 1 \,.
\eal

\subsection{Parametric-resonance approach}

Note that
\bal
\hat{J}_+ =  \hat{a}^\dagger \hat{c}^\dagger \,, \quad \hat{J}_- = \hat{c} \hat{a} \,, \quad \hat{J}_3 = \frac{1}{2} \left(\hat{a}^\dagger \hat{a} - \hat{c} \hat{c}^\dagger\right)
\eal
satisfy the Lie algebra of SU(2):
\bal
[\hat{J}_+, \hat{J}_-] = 2 \hat{J}_3 \,, \quad [\hat{J_3}, \hat{J}_\pm] = \pm \hat{J}_\pm \,.
\eal
In standard textbooks (e.g., Ref.~\cite{Sakurai:2011zz}), this problem is often formulated in a $2 \times 2$ matrix form by taking the following representations with Pauli matrices:
\bal
\hat{J}_\pm = \sigma_\pm = \frac{1}{2} \left( \sigma_1 \pm i \sigma_{2} \right) \,, \quad \hat{J}_3 = \frac{1}{2} \sigma_3 \,.
\eal
Though it might be easier to deal with the $2 \times 2$ matrix, it would be still illustrative deriving the result in a similar way to the Bosonic model as follows. 

The time-evolution (Unitary) operator is given by
\bal
&\hat{U}(t) = \hat{U}_{0}(t) \hat{U}_{I}(t) \,, \quad 
\hat{U}_{0}(t) = \exp\left[ - i \frac{m}{2} t \left(\hat{a}^\dagger \hat{a} - \hat{c} \hat{c}^\dagger\right) \right] \,, \notag \\
&\hat{U}_{I}(t) = \exp\left[ - i \left( \omega - \frac{m}{2} \right) t \left(\hat{a}^\dagger \hat{a} - \hat{c} \hat{c}^\dagger\right) - i \gamma t e^{- i \phi} \hat{c} \hat{a} - i \gamma t e^{i \phi} \hat{a}^\dagger \hat{c}^\dagger \right] \,.
\eal
One can check that it satisfies
\bal
i \frac{d}{dt} \hat{U}(t) = \hat{H} (t) \hat{U}(t) \,,
\eal
by using
\bal
&\hat{U}_0(t) \hat{a} \hat{U}^\dagger_0(t) = e^{i m t / 2} \hat{a} \,, \quad \hat{U}_0(t) \hat{a}^\dagger \hat{U}^\dagger_0(t) = e^{- i m t / 2} \hat{a}^\dagger \,, \notag \\
&\hat{U}_0(t) \hat{c} \hat{U}^\dagger_0(t) = e^{i m t / 2} \hat{c} \,, \quad \hat{U}_0(t) \hat{c}^\dagger \hat{U}^\dagger_0(t) = e^{- i m t / 2} \hat{c}^\dagger \,.
\eal
We consider the following time-dependent annihilation and creation operators:
\bal
& \hat{b}(t) = \hat{U}_I(t) \hat{a} \hat{U}_I^\dagger(t) \,, \quad \hat{b}^\dagger(t) = \hat{U}_I(t) \hat{a}^\dagger \hat{U}_I^\dagger(t) \,, \notag \\
& \hat{d}(t) = \hat{U}_I(t) \hat{c} \hat{U}_I^\dagger(t) \,, \quad \hat{d}^\dagger(t) = \hat{U}_I(t) \hat{c}^\dagger \hat{U}_I^\dagger(t) \,,
\eal
which follow
\bal
&i \frac{d}{dt} \hat{b}(t) = - \left( \omega - \frac{m}{2} \right) \hat{b}(t) - \gamma e^{i \phi} \hat{d}^\dagger(t) \,, \quad
i \frac{d}{dt} \hat{b}^\dagger(t) = \left( \omega - \frac{m}{2} \right) \hat{b}^\dagger(t) + \gamma e^{- i \phi} \hat{d}(t) \,, \notag \\
&i \frac{d}{dt} \hat{d}(t) = \gamma e^{i \phi} \hat{b}^\dagger(t) - \left( \omega - \frac{m}{2} \right) \hat{d}(t) \,, \quad
i \frac{d}{dt} \hat{d}^\dagger(t) = - \gamma e^{- i \phi} \hat{b}(t) + \left( \omega - \frac{m}{2} \right) \hat{d}^\dagger(t) \,.
\eal
One can check that the followings satisfy these equations:
\bal
& \hat{b}(t) = \left[ \cos(\Omega t) + i \frac{\sin(\Omega t)}{\Omega} \left( \omega - \frac{m}{2} \right) \right] \hat{a} + i \frac{\sin(\Omega t)}{\Omega} \gamma e^{i \phi}  \hat{c}^\dagger \,, \notag \\
& \hat{b}^\dagger(t) = \left[ \cos(\Omega t) - i \frac{\sin(\Omega t)}{\Omega} \left( \omega - \frac{m}{2} \right) \right] \hat{a}^\dagger - i \frac{\sin(\Omega t)}{\Omega} \gamma e^{- i \phi} \hat{c} \,, \notag \\
& \hat{d}(t) = - i \frac{\sin(\Omega t)}{\Omega} \gamma e^{i \phi}  \hat{a}^\dagger + \left[ \cos(\Omega t) + i \frac{\sin(\Omega t)}{\Omega} \left( \omega - \frac{m}{2} \right) \right] \hat{c}  \,, \notag \\
& \hat{d}^\dagger(t) = i \frac{\sin(\Omega t)}{\Omega} \gamma e^{- i \phi} \hat{a} + \left[ \cos(\Omega t) - i \frac{\sin(\Omega t)}{\Omega} \left( \omega - \frac{m}{2} \right) \right] \hat{c}^\dagger \,,
\eal
which are even functions of $\Omega$ with
\bal
\Omega^2 = \left( \omega - \frac{m}{2} \right)^2 + \gamma^2 \,.
\eal
As a result, one obtains
\bal
&\hat{a}(t) = \hat{U}(t) \hat{a} \hat{U}^\dagger(t) = \mu(t) \hat{a} + \nu(t) \hat{c}^\dagger \,, \quad \hat{a}^\dagger(t) = \hat{U}(t) \hat{a}^\dagger \hat{U}^\dagger(t) = \mu^{*}(t) \hat{a}^\dagger + \nu^{*}(t) \hat{c} \notag \\
&\hat{c}(t) = \hat{U}(t) \hat{c} \hat{U}^\dagger(t) = - \nu(t) \hat{a}^\dagger + \mu(t) \hat{c} \,, \quad \hat{c}^\dagger(t) = \hat{U}(t) \hat{c}^\dagger \hat{U}^\dagger(t) = - \nu^{*}(t) \hat{a} + \mu^{*}(t) \hat{c}^\dagger  \notag \\
& \mu(t) = \left[ \cos(\Omega t) + i \frac{\sin(\Omega t)}{\Omega} \left( \omega - \frac{m}{2} \right) \right] e^{i m t / 2} \,, \quad \nu(t) = i \frac{\sin(\Omega t)}{\Omega} \gamma e^{i \phi} e^{- i m t / 2} \,.
\eal
One can check that
\bal
|\mu(t)|^2 + |\nu(t)|^2 = 1 \,,
\eal
and thus the inverse relation is given by
\bal
&\hat{a}_H(t) = \hat{U}^\dagger(t) \hat{a} \hat{U}(t) = \mu^*(t) \hat{a} - \nu(t) \hat{c}^\dagger \,, \quad \hat{a}^\dagger_H(t) = \hat{U}^\dagger(t) \hat{a}^\dagger \hat{U}(t) = \mu(t) \hat{a}^\dagger - \nu^{*}(t) \hat{c} \,, \notag \\
&\hat{c}_H(t) = \hat{U}^\dagger(t) \hat{c} \hat{U}(t) = \nu(t) \hat{a}^\dagger + \mu^*(t) \hat{c}  \,, \quad \hat{c}^\dagger_H(t) = \hat{U}^\dagger(t) \hat{c}^\dagger \hat{U}(t) = \nu^{*}(t) \hat{a} + \mu(t) \hat{c}^\dagger \,.
\eal
These relations are called the Bogoliubov transformation.

The state evolves from the vacuum,
\bal
|0\rangle (t) = \hat{U}(t) |0\rangle \,,
\eal
satisfies
\bal
\hat{a}(t) |0\rangle (t) = \hat{c}(t) |0\rangle (t) = 0 \,.
\eal
One can check that it is given by the number operators,
\bal
| m, n \rangle = \hat{a}^{\dagger m} \hat{c}^{\dagger n} | 0 \rangle \,,
\eal
as (up to the irrelevant overall phase $\alpha(t)$)
\bal
|0\rangle (t) = e^{i \alpha(t)} \sqrt{1 - |\nu(t)|^2} \left( |0, 0 \rangle - \frac{\nu (t)}{\mu (t)} | 1, 1 \rangle \right)\,.
\eal
Therefore, the vacuum-to-vacuum transition amplitude with the time interval $T$ is given by
\bal
|_{\rm out}\langle 0 | 0 \rangle_{\rm in}| = \sqrt{1 - |\nu(T)|^2} = \sqrt{1 - \gamma^2 \sin^2 (\Omega T) / \Omega^2} \,.
\eal
Here note that the expectation number of produced particles is given by 
\bal
f(t) = \langle 0| \hat{a}^\dagger_H(t) \hat{a}_H(t) | 0 \rangle = \langle 0| \hat{c}^\dagger_H(t) \hat{c}_H(t) | 0 \rangle = |\nu(t)|^{2} \,,
\eal
and thus $|_{\rm out}\langle 0 | 0 \rangle_{\rm in}| = \sqrt{1 - f(T)}$.
We find (possibly time-dependent) decay rate as
\bal
\Gamma^P_\omega(T) = - \frac{1}{T} \ln \left( 1 - \gamma^2 \sin^2 (\Omega T) / \Omega^2 \right) \,.
\eal

\subsection{Feynman-diagrammatic approach}
The Feynman-diagrammatic approach goes in a very similar way to the Bosonic model.
The propagator is given by
\bal
&\langle 0 | T\left\{ \hat{a}(t_1) \hat{a}^\dagger(t_2) \right\} | 0 \rangle = - \langle 0 | T\left\{ \hat{a}^\dagger(t_2) \hat{a}(t_1)  \right\} | 0 \rangle = \langle 0 | T\left\{ \hat{b}(t_1) \hat{b}^\dagger(t_2) \right\} | 0 \rangle = - \langle 0 | T\left\{ \hat{b}^\dagger(t_2) \hat{b}(t_1) \right\} | 0 \rangle \notag \\
&= \theta(t_1 - t_2) e^{- i \omega (t_1 - t_2)}  = \int \frac{d \omega'}{2 \pi} \frac{i}{\omega' - \omega + i \epsilon} e^{- i \omega' (t_1 - t_2)} \,.
\eal
Note the minus sign, arising from the Fermionic nature of the annihilation/creation operators, in the first line.
The amplitude with $p$-pair insertion is given by
\bal
i {\cal M}_{p} = \frac{(-1)^{p-1}}{p}\gamma^{2p} \int \frac{d \omega'}{2 \pi} (\omega' - \omega + i \epsilon)^{-p} (m - \omega' - \omega + i \epsilon)^{-p} \,,
\eal
where the prefactor $1/p$ is a symmetric factor and $(-1)^{p-1}$ arises from the the Fermionic nature.
One obtains
\bal
\Gamma^F_\omega = \sum_{p=1}^\infty 2 \pi \frac{(-1)^{p-1}}{p}\gamma^{2p} \frac{1}{[(p-1)!]^2} \lim_{\omega_1\to \omega} \lim_{\omega_2\to \omega} \frac{\partial^{p-1}}{\partial \omega_1^{p-1}} \frac{\partial^{p-1}}{\partial \omega_2^{p-1}} \delta(m - \omega_1 - \omega_2) \,.
\eal

\subsection{Comparison of two approaches}
As in the Bosonic Rabi model, we consider the integral with a test function $f(w)$.
In the Feynman-diagrammatic approach, one obtains
\bal
\int dw f(w) \Gamma^F_{p, w} = 4 \pi (-1)^{p-1} \gamma^{2p} \frac{1}{p!(p-1)!} \frac{1}{2^{2p}} \frac{d^{2p-2}}{d w^{2p-2}} f(w=0) \,.
\eal
In the parametric-resonance approach, one expects the following Laurent expansion of $\Gamma^{P, +}_{p, w}$ around $w = 0$ [in the course of the same discussion around \eqref{eq:pole}]:
\bal
\Gamma^{P, +}_{p, w} = - i (-1)^p \frac{(2p-3)!!}{2^p p!} \frac{1}{w^{2p-1}}  \gamma^{2p}+ {\cal O}(w^0)\gamma^{2p}.
\eal
The leading 3 orders are given by the Bosonic result multiplied by $2 (-1)^p$ to confirm the expectation.
One can see $\int d w f(w) \Gamma^P_\omega = \int d w f(w) \Gamma^F_\omega$ at all orders of $\gamma^2$.

We also examine the non-perturbative integral in the parametric-resonance approach:
\bal
\int dw f(w) \Gamma^P_w = - \frac{1}{T} \int dw f(w) \ln \left[ 1 - \gamma^2 \sin^2 (\sqrt{w^2 + \gamma^2} T) / (w^2 + \gamma^2) \right] \,.
\eal
Note that there is no growing mode and thus no obvious asymptotic limit. One can rewrite the integral with $w = \gamma x $ and $\tau = \gamma T$:
\bal
\int dw f(w) \Gamma^P_w = - \frac{\gamma^2}{\tau} \int dx f(\gamma x) \ln \left[ 1 - \sin^2 (\sqrt{x^{2}+1} \tau) / (x^2 + 1) \right] \,,
\eal

First, we consider $f(w) = 1$.
The numerical-integration result in the parametric-resonance approach is presented in Fig.~\ref{fig:Rabi1} and compared with the Feynman-diagrammatic approach (actually, one can perform the integration analytically; see appendix~\ref{sec:integral}):
\bal
\int dw \Gamma^F_w = \pi \gamma^2 \,.
\eal
Second, we consider $f(w)=w_0^2 / (w_0^2 + w^2)$ (we take the units of $w_0 = 1$ in numerical computations).
The numerical-integration result is presented in Fig.~\ref{fig:Rabi2} and compared with the Feynman-diagrammatic approach:
\bal
\int dw \frac{w_0^2}{w_0^2 + w^2} \Gamma^F_w = 2 \pi \gamma^2 \frac{1}{\sqrt{1 - (\gamma/w_0)^2} + 1} \,.
\eal

\section{Integral in the Rabi models} \label{sec:integral}
First, we consider the following integral in the Bosonic Rabi model:
\bal
I = \int dz \ln \left[ 1 + \sin^2 (\sqrt{z^{2}-1} \tau) / (z^2 - 1) \right] \,.
\eal
One can rewrite it as
\bal
& I = 2 {\rm Re} \int dz \ln \left[ f(z) \right] \,, \notag \\
& f(z) = \cos (\sqrt{z^2-1} \tau) - i z \sin (\sqrt{z^2-1} \tau) / \sqrt{z^2-1} \notag \\
& \qquad = \frac{1}{2} \left[- \left(z / \sqrt{z^2-1} - 1\right)  e^{i \sqrt{z^2-1} \tau} + \left(z / \sqrt{z^2-1} + 1\right) e^{-i \sqrt{z^2-1} \tau} \right] \,.
\eal
Since $\cos (\sqrt{z^2-1} \tau)$ and $\sin (\sqrt{z^2-1} \tau) / \sqrt{z^2-1}$ are an even function of $\sqrt{z^2-1}$, $f(z)$ is analytic everywhere in the complex $z$ plane.
Therefore, any non-trivial complex structure of $\ln f(z)$ can only come from the branch cut extending from a zero of $f(z)$.


Any zero of $f(z)$ satisfies
\bal
e^{2 i \sqrt{z^2-1} \tau} = (z + \sqrt{z^2-1})/(z - \sqrt{z^2-1}) \,.
\eal
Taking the squared absolute value of the both sides, one finds
\bal
e^{- 4 {\rm Im} (\sqrt{z^2-1}) \tau} = | z + \sqrt{z^2-1} |^2/ | z - \sqrt{z^2-1} |^2 \,.
\eal
When $z$ is in the first (second) quadrant, we take $\sqrt{z^2-1}$ so that it is also in the first (second) quadrant.
The left hand side is smaller than or equal to unity in the upper half plane.
Concerning the right hand side, $| z + \sqrt{z^2-1} |^2 - | z - \sqrt{z^2-1} |^2 = 4 {\rm Re} (z^* \sqrt{z^2-1}) \geq 0$ in the upper half plane; namely, the right hand side is bigger than or equal to unity.
Therefore, any zero in the upper half plane makes both the left and right hand sides equal to unity.
Concerning the left hand side, it corresponds to $z \leq -1$ or $z \geq 1$.
Concerning the right hand side, it corresponds to $-1 \leq z \leq 1$.
Therefore, only $z = \pm 1$ are candidates, but one can easily see that they are not zeros.
Therefore, there is no zero of $f(z)$ in the upper half plane.

Because of the analyticity of the integrand, one can change the integration contour to a large clockwise semicircle in the upper half plane (C):
\bal
I = - 2 {\rm Re} \int_C dz \ln \left[ e^{i z \tau} f(z) \right] \,.
\eal
Here we multiply $e^{i z \tau}$ by $f(z)$, which does not change the real part of the integral along real $z$ nor introduce branch cut of logarithm.
Since for a large $z$ in the upper half plane,
\bal
e^{i z \tau} f(z) \approx e^{i (z - \sqrt{z^2-1}) \tau} \approx  e^{i  \tau / (2 z)} \,,
\eal
\bal
I = - 2 {\rm Re} \left[ i \pi (i \tau / 2) \right] = \pi \tau  \,.
\eal

Second, we consider the following integral in the Fermionic Rabi model:
\bal
I = - \int dz \ln \left[ 1 - \sin^2 (\sqrt{z^{2}+1} \tau) / (z^2 + 1) \right] \,.
\eal
The following steps are similar to the Bosonic model and thus omit the repetitive discussions to highlight the difference.
One can rewrite it as
\bal
& I = - 2 {\rm Re} \int dz \ln \left[ f(z) \right] \,, \notag \\
& f(z) = \cos (\sqrt{z^2+1} \tau) - i z \sin (\sqrt{z^2+1} \tau) / \sqrt{z^2+1} \notag \\
& \qquad = \frac{1}{2} \left[\left(1 - z / \sqrt{z^2+1} \right)  e^{i \sqrt{z^2+1} \tau} + \left(1 + z / \sqrt{z^2+1}\right) e^{-i \sqrt{z^2+1} \tau} \right] \,.
\eal

Any zero of $f(z)$ satisfies
\bal
e^{2 i \sqrt{z^2+1} \tau} = - (\sqrt{z^2+1} + z )/(\sqrt{z^2+1} - z) \,.
\eal
Taking the squared absolute value of the both sides, one finds
\bal
e^{- 4 {\rm Im} (\sqrt{z^2+1}) \tau} = | \sqrt{z^2+1} + z |^2/ | \sqrt{z^2+1} - z |^2 \,.
\eal
When $z$ is in the first (second) quadrant, we take $\sqrt{z^2+1}$ so that it is also in the first (second) quadrant.
The left hand side is smaller than or equal to unity in the upper half plane.
Concerning the right hand side, $| \sqrt{z^2+1} + z |^2 - | \sqrt{z^2+1} - z |^2 = 4 {\rm Re} (z^* \sqrt{z^2+1}) \geq 0$ in the upper half plane; namely, the right hand side is bigger than or equal to unity.
Therefore, any zero in the upper half plane makes both the left and right hand sides equal to unity.
Concerning the left hand side, it corresponds to real z or imaginary $z = i y$ with $0 \leq y \leq 1$.
Concerning the right hand side, it corresponds to imaginary $z = i y$ with $0 \leq y \leq 1$.

For $z = i y$ with $0 \leq y \leq 1$,
\bal
f(z) = \cos (\sqrt{1-y^2} \tau) + y \sin (\sqrt{1-y^2} \tau) / \sqrt{1-y^2} \,.
\eal
Substituting $y = \sin (\theta)$ with $0 \leq \theta < \pi/2$ (omitting $y=1$; one can see that is it is not a zero), one can find
\bal
f(z) = \cos [\cos (\theta) \tau] + \sin (\theta) \sin [\cos (\theta) \tau] / \cos (\theta) = \cos [\cos (\theta) \tau - \theta] / \cos (\theta)\,.
\eal
Zeros are given by
\bal
\cos (\theta) \tau - \theta = (n - 1/2) \pi \,,
\eal
with integer $n$.
The left hand side is monotonically decreasing function from $\tau$ to $- \pi/2$.
Therefore, there are $\lfloor\tau / \pi + 1/2\rfloor$ ($\lfloor x\rfloor$ is the floor function) zeros.
There is no zero for $\tau < \pi/2$.

Therefore, we multiply not only $g(z) = \prod_n [z + i y_n (\tau)] / [z - i y_n (\tau)]$ as well as $e^{i z \tau}$ by $f(z)$ to remove branch cut of logarithm from the upper half plane.
It does not change the real part of the integral along real $z$.
Because of the analyticity of the integrand, one can change the integration contour to a large clockwise semicircle in the upper half plane (C):
\bal
I = 2 {\rm Re} \int_C dz \ln \left[ e^{i z \tau} g(z) f(z) \right] \,.
\eal
Since for a large $z$ in the upper half plane,
\bal
e^{i z \tau} g(z) f(z) \approx e^{i (z - \sqrt{z^2+1}) \tau} \prod_n [1 + i y_n (\tau) / z]/[1 - i y_n (\tau) / z] \approx  e^{-i  \tau / (2 z)} \left(1 + 2 i \left[\sum_ n y_n (\tau) \right] / z \right) \,,
\eal
one finds
\bal
I = 2 {\rm Re} \left( i \pi \left[-i \tau / 2 + 2 i \sum_ n y_n (\tau) \right] \right) = \pi \tau - 4 \pi \sum_ n y_n (\tau) \,.
\eal

\acknowledgments
We acknowledge Kai Murai met Progress in Particles Physics 2026 (YITP-W-26-9) and ChatGPT (OpenAI) for assistance with integrals in appendix~\ref{sec:integral}.
K. S. acknowledges Gemini (Google) for assistance with development of numerical codes.
All scientific content, analysis, and conclusions are the sole responsibility of the authors.
A. K. acknowledges partial support from Norwegian Financial Mechanism for years 2014-2021, grant nr 2019/34/H/ST2/00707; and from National Science Centre, Poland, grant DEC-2018/31/B/ST2/02283.
K.S. acknowledges support by Japan Society for the Promotion of Science (JSPS) under grant 25K17379 and 21K20363 and Tsuruoka National College of Technology Promotion Association under grant.

\bibliography{biblio} 
\bibliographystyle{JHEP}

\end{document}